\documentclass[aps,prl,reprint,superscriptaddress]{revtex4-1}

\usepackage{amsmath}
\usepackage[amssymb]{SIunits}
\usepackage{url}
\usepackage{graphicx}
\usepackage{color}
\usepackage{soul}

\usepackage{hyperref}
\usepackage{mathtools}

\begin{document}

\title{Semiconductor ring laser frequency combs induced by phase turbulence}
	
\author{Marco Piccardo$^\dagger$}
\affiliation{Harvard John A. Paulson School of Engineering and Applied Sciences, Harvard University, Cambridge, MA 02138, USA}

\author{Benedikt Schwarz$^\dagger$}
\affiliation{Harvard John A. Paulson School of Engineering and Applied Sciences, Harvard University, Cambridge, MA 02138, USA}
\affiliation{Institute of Solid State Electronics, TU Wien, 1040 Vienna, Austria}

\author{Dmitry Kazakov}
\affiliation{Harvard John A. Paulson School of Engineering and Applied Sciences, Harvard University, Cambridge, MA 02138, USA}

\author{Maximilian Beiser}
\affiliation{Institute of Solid State Electronics, TU Wien, 1040 Vienna, Austria}

\author{Nikola Opacak}
\affiliation{Institute of Solid State Electronics, TU Wien, 1040 Vienna, Austria}

\author{Yongrui Wang}
\affiliation{Department of Physics and Astronomy, Texas A\&M University, College Station, TX 77843, USA}

\author{Shantanu Jha}
\affiliation{Harvard John A. Paulson School of Engineering and Applied Sciences, Harvard University, Cambridge, MA 02138, USA}
\affiliation{Physics Department, Yale University, New Haven, CT 06511, USA}

\author{Michele Tamagnone}
\affiliation{Harvard John A. Paulson School of Engineering and Applied Sciences, Harvard University, Cambridge, MA 02138, USA}

\author{Wei Ting Chen}
\affiliation{Harvard John A. Paulson School of Engineering and Applied Sciences, Harvard University, Cambridge, MA 02138, USA}

\author{Alexander Y. Zhu}
\affiliation{Harvard John A. Paulson School of Engineering and Applied Sciences, Harvard University, Cambridge, MA 02138, USA}

\author{Lorenzo L. Columbo}
\affiliation{Dipartimento di Elettronica e Telecomunicazioni, Politecnico di Torino, Corso Duca degli Abruzzi 24, 10129 Torino, Italy}

\author{Alexey Belyanin}
\affiliation{Department of Physics and Astronomy, Texas A\&M University, College Station, TX 77843, USA}

\author{Federico Capasso}
\email[]{capasso@seas.harvard.edu}
\affiliation{Harvard John A. Paulson School of Engineering and Applied Sciences, Harvard University, Cambridge, MA 02138, USA}

\collaboration{$^\dagger$These authors contributed equally to this work.}

\begin{abstract}
Semiconductor ring lasers are miniaturized devices that operate on clockwise and counterclockwise modes. These modes are not coupled in the absence of intracavity reflectors, which prevents the formation of a standing wave in the cavity and, consequently, of a population inversion grating. This should inhibit the onset of multimode emission driven by spatial hole burning. Here we show that, despite this notion, ring quantum cascade lasers inherently operate in phase-locked multimode states, that switch on and off as the pumping level is progressively increased. By rewriting the master equation of lasers with fast gain media in the form of the complex Ginzburg-Landau equation, we show that ring frequency combs stem from a phase instability---a phenomenon also known in  superconductors and Bose-Einstein condensates. The instability is due to coupling of the amplitude and phase modulation of the optical field in a semiconductor laser, which plays the role of a Kerr nonlinearity, highlighting a connection between laser and microresonator frequency combs.
\end{abstract}

\maketitle

\begin{figure*}[t!]
    \centering
    \includegraphics[width=1\textwidth]{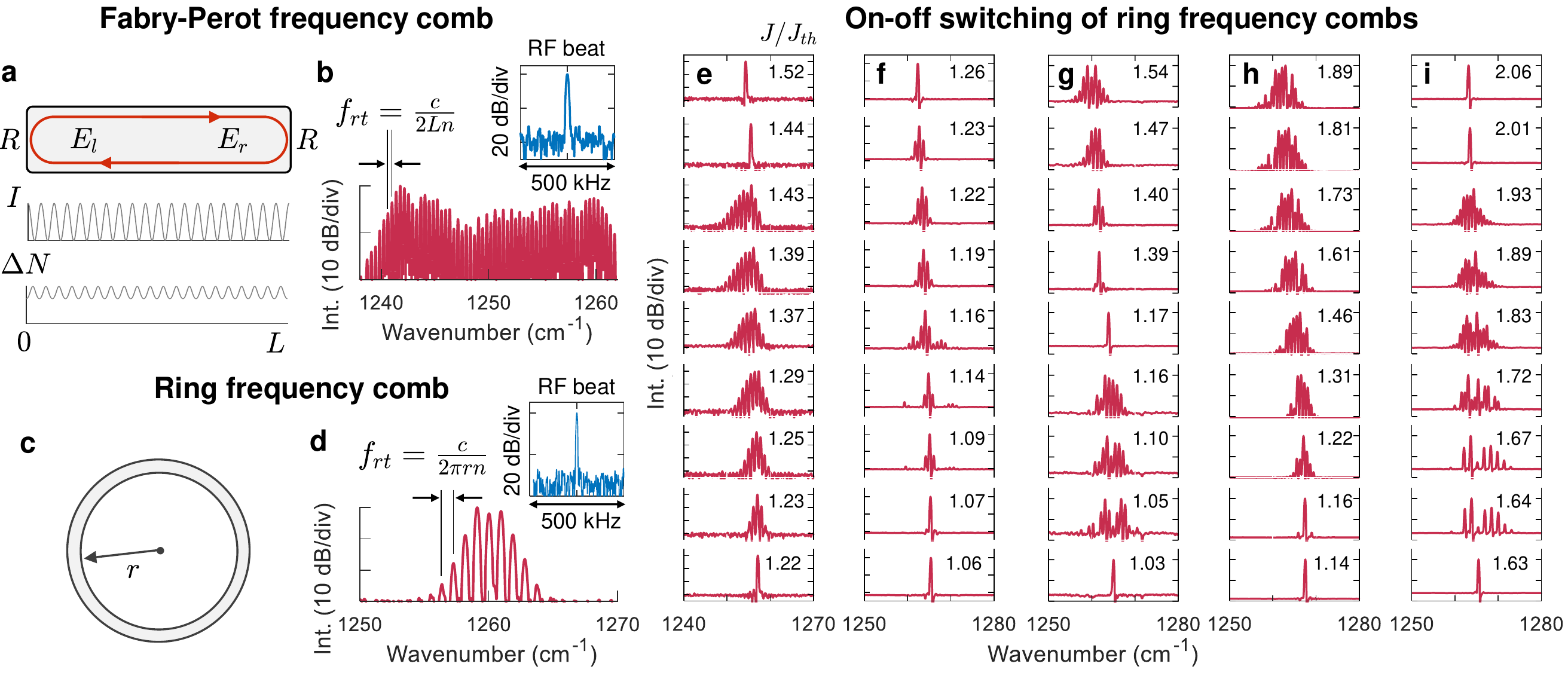}
    \caption{\textbf{a}, Schematic of a Fabry-Perot cavity with rightward and leftward propagating waves, $E_r$ and $E_l$, that are coupled through the facets reflectivity $R$. As a result of counter-propagating waves, optical standing waves of intensity $I$ and a static grating of population inversion $\Delta N$ are formed in the cavity of length $L$. \textbf{b}, Optical spectrum of a Fabry-Perot frequency comb generated from a quantum cascade laser ($L=3.7$~mm). The roundtrip frequency ($f_{rt}$) defines the comb spacing and the beat note frequency ($\approx$6.7~GHz), whose spectrum is shown in the inset. \textbf{c}, Schematic of a semiconductor ring laser of radius $r$. \textbf{d}, Optical spectrum of a ring quantum cascade laser ($r=500$~$\mu$m) measured at an injection current density fractionally higher than the lasing threshold. Also shown is the narrow electrical beatnote of the laser (central frequency $\approx$27.8~GHz). \textbf{e-i}, Evolution of the optical spectra of five distinct ring lasers with increasing injected current showing that the multimode regime can switch on and off. The current density normalized to the lasing threshold is given to the right of each spectrum.}
    \label{fig_perfectring}
\end{figure*}

Despite decades of studies, the development of optical frequency combs continues at a rapid pace~\cite{Picque2019,Gaeta2019,Kues2019}. While historically research on frequency combs started from tabletop optical systems, such as Ti:sapphire modelocked lasers, that revolutionized optical metrology~\cite{Hansch2006}, the following advances in semiconductor and dielectric materials processing led to much more compact frequency comb generators. Technological progress went hand in hand with a burst of new applications, such as broadband spectroscopy and chemical sensing, radiofrequency arbitrary waveform generation, optical communications and quantum information.

Within the realm of integrated optics of particular interest are two classes of generators: semiconductor lasers, which embed an active medium that is internally pumped, and passive microresonators~\cite{Gaeta2019,Kippenberg555}, where the gain stems from the optical Kerr nonlinearity and the pump is an external continuous-wave laser. Whether in the case of a semiconductor laser or of a microresonator, the device always starts from single-frequency operation, corresponding to the first lasing mode or the external pump. Therefore, to trigger generation of a frequency comb, mechanisms capable of coupling modes at different frequencies and locking their phases need to be present in the cavity. In microcavity resonators above the parametric instability threshold, an external pump induces the appearance of sidebands, which proliferate, upon amplification, by means of cascaded parametric processes~\cite{Herr2012}. In semiconductor lasers multimode operation is typically induced by the standing wave created by the first lasing mode (Fig.~\ref{fig_perfectring}a), which leads by means of stimulated emission to a spatially inhomogeneous distribution of the gain---a phenomenon known as spatial hole burning (SHB). Phase locking is achieved by means of an optical nonlinearity, such as saturable absorption~\cite{Kaertner1995} or four-wave-mixing~\cite{Agrawal1988,Faist2016}. SHB is not expected to occur in a ring cavity, as the clockwise (CW) and counter-clockwise (CCW) modes are not naturally coupled in absence of any reflection points in the cavity. We show that semiconductor ring lasers can operate in multimode regimes and form frequency combs~\cite{Gelens2009} in absence of SHB. Multimode emission and comb formation occurs  at a pumping level fractionally higher than the lasing threshold, thus excluding the Risken-Nummedal-Graham-Haken instability, that also promotes multimode operation of a laser, as it is predicted to occur at a pumping level of nine times above threshold~\cite{Risken1968,Graham1968}. We explain this behaviour of a ring laser by showing that it can be described by the Ginzburg-Landau theory~\cite{Aranson2002}. It predicts that a semiconductor laser can transition to a multimode regime at low pumping due to coupling between intensity and phase noise. This coupling is inherent to any semiconductor laser and is quantified by the linewidth enhancement factor (LEF).


We study ring quantum cascade lasers~\cite{Mujagic2008} (QCLs)---monolithic frequency comb generators that combine nonlinearity and gain~\cite{Hugi2012} and are targeting applications in dual-comb spectroscopy~\cite{Villares2014}, metrology~\cite{consolino2019fully} and microwave photonics~\cite{PiccardoLRT}. Several works investigated QCL cavities with circular geometry. Monolithic rings with distributed-feedback~\cite{Mujagic2008} and metamaterial~\cite{Szedlak2018} gratings were studied for surface outcoupling of single modes. External ring cavities were demonstrated for active mode-locking applications~\cite{Malara2013,Wojcik2013,Revin2016}. Additionally, microcavities, such as disks~\cite{Faist1996}, bow-tie~\cite{Gmachl1556} and elliptical resonators~\cite{Wang22407} were investigated. However, none of these works studied generation of self-starting frequency combs in lasers with such cavity arrangements. In this work we fabricated ring QCLs in a ridge waveguide geometry with a width of 10~$\mu$m and an inner radius $r$ of 500 or 600 $\mu$m (Fig.~\ref{fig_perfectring}c). The active region consists of AlInAs/GaInAs/InP layers and the band structure design is based on a single-phonon continuum depopulation scheme~\cite{WangQCLmaterial2017}. Emission is at 7.9~$\mu$m and operation is under constant electrical injection at room temperature.

\begin{figure*}[t!]
    \centering
    \includegraphics[width=0.95\textwidth]{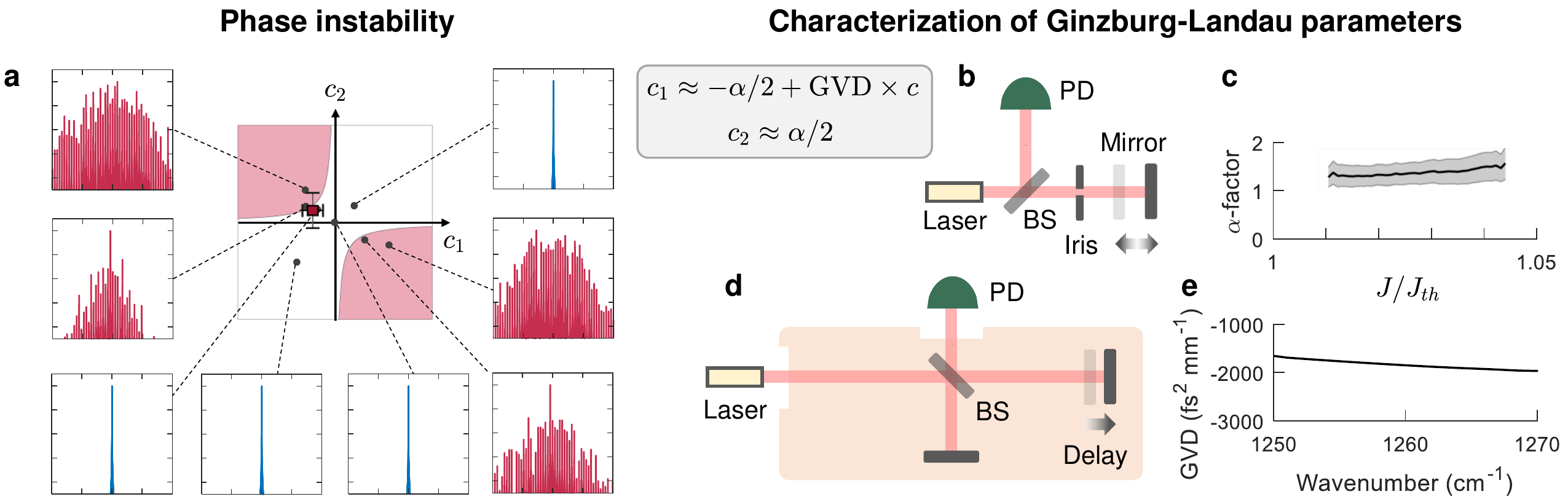}
    \caption{\textbf{a} We show that ring lasers with fast gain media obey the complex Ginzburg-Landau equation, which is governed by only two parameters, $c_1$ and $c_2$. In the Ginzburg-Landau theory the parameter space can be divided into three regions: a central one (white area) where the running wave admits stable single mode solutions, and two outer ones (red areas) where the wave possesses multimode solutions. Spectra obtained from space-time domain simulations of the laser master equation are shown for different points in parameter space confirming the behavior expected from the Ginzburg-Landau theory. In the simulations spatial hole burning is turned off, thus the multimode regimes, where observed, are due to the Ginzburg-Landau phase instability. In all plots the x-axis spans 100 longitudinal modes of the ring, the y-axis is intensity (10 dB/div). Also shown is the region corresponding to the experimentally studied devices as obtained from the laser parameters with related uncertainties (square marker with error bars). \textbf{b-e}, Characterization of the physical quantities that define $c_1$ and $c_2$: $\alpha$ is the linewidth enhancement factor, GVD is the group velocity dispersion, $c$ is a constant (Supplementary Material). \textbf{b}, Self-interferometry setup used for the measurement of the $\alpha$-factor. \textbf{c}, The experimental values are shown as a function of current density normalized to the lasing threshold. \textbf{d}, Schematic of the setup used to measure the GVD and \textbf{e}, corresponding measurement shown for the emission range of the ring lasers studied here.}
    \label{fig_phaseturbulence}
\end{figure*}

A small fraction of light can escape from the ring cavity, due to scattering induced by waveguide roughness and bending losses, allowing us to perform their spectral characterization. Our finding is that at a current injection level that is only fractionally higher than the lasing threshold (typically $1.1-1.2~J_{th}$) the ring laser undergoes a transition to a multimode regime that is characteristically different from the one observed in Fabry-Perot QCLs (Fig.~\ref{fig_perfectring}b). The optical spectrum has fewer modes and exhibits a regular, bell-shaped envelope (Fig.~\ref{fig_perfectring}d). 
The modes are separated by the roundtrip frequency of the laser $f_{rt}=c/(2\pi r n)$, where $n=3.4$ is the effective refractive index of the waveguide. The coherence of the state is witnessed by its narrow beat note, which proves its frequency comb nature. Spectral evolution of five representative devices is shown in Fig.~\ref{fig_perfectring} e-i. A general feature of the instability is that as the current in the device is increased, the multimode regime switches on and off, as the laser reverts back to single mode operation at high current. Movement of the center of mass of the spectrum is smooth with current and shows no discontinuous jumps, which precludes the argument that such behaviour could be caused by other destabilization mechanisms---electrical, mechanical or thermal. Such spectral evolution is not observed in regular Fabry-Perot cavity lasers: there the spectrum broadens as the injection current is progressively increased and the laser never reverts to single-mode operation once the multimode regime is reached. 

To support the experimental evidence of multimode operation we reexamine the theory of lasers with fast gain media. We will show that ring frequency combs can be explained on the basis of a phase instability that affects the single mode solution of the complex Ginzburg-Landau equation (CGLE). CGLE is a nonlinear equation that describes spatially-extended systems of coupled nonlinear oscillators. It appears in many branches of physics, such as superconductivity, Bose-Einstein condensation and quantum field theory. In semiconductor laser theory it can be shown that the field dynamics can be described by a CGLE in the hypothesis of fast gain~\cite{Gil2014}. While the latter is not suitable for a conventional bipolar semiconductor laser (diode laser) it is instead very well verified for QCLs. Here we rewrite the master equation of lasers with fast gain media~\cite{opacak2019theory}, such as QCLs, in the CGLE form as
\begin{equation}
    \partial_t E  = E + (1+i c_1) \partial^2_z E - (1 + i c_2) |E|^2 E
\end{equation}
where $E$ is the electric field, $t$ is time, $z$ is the spatial coordinate running along the ring cavity (see the Supplementary Material for the analytical derivation). The only two parameters of the equation are $c_1$ and $c_2$, which capture the stability of the system. In the case of QCLs, $c_1$ is directly proportional to the group velocity dispersion (GVD), while $c_2$ depends on the Kerr coefficient. The latter is normally small in QCLs but its effect is compensated by the LEF, which can be effectively regarded as a Kerr nonlinearity (Supplementary Material). The LEF appears in both $c_1$ and $c_2$, and provides phase-amplitude coupling needed for phase-locking of multiple oscillating modes. It has an effect on the gain profile making it asymmetric and providing a carrier-dependent contribution to the real part of the complex refractive index~\cite{opacak2019theory}. 

In CGLE theory, the parameter space is divided into different regions by the Benjamin-Feir lines~\cite{Aranson2002,Chate1994}, which are defined by $1 + c_1 c_2 =0$. The inner region confined by the lines has stable, purely single-mode solutions, while the two outer regions exhibit a so-called phase instability~\cite{Gil2014}, i.e., they possess multimode solutions. In Fig.~\ref{fig_phaseturbulence}f we show the result of space-time domain simulations of a ring QCL for different points in the $(c_1,c_2)$-parameter space determined by laser parameters with typical values. We note that in these numerical simulations no approximation is made to reduce the laser equations that capture the whole dynamics of the system to the CGLE. The computed optical spectra confirm that in the stability region only single mode solutions are supported, while in the outer regions the laser can attain a multimode regime despite the absence of SHB, as already suggested by a recent theory of ring QCLs based on effective Maxwell-Bloch equations~\cite{Columbo2018}. In CGLE theory one would refer to the dynamical behaviour emerging from the phase instability of QCLs as ``phase turbulence". This is because the amplitude of the field is almost constant, while its phase fluctuates in time, as we observe in the numerical simulations (not shown here).

In order to connect the phase instability with the investigated ring devices, we measured both GVD and LEF of our laser material. The LEF was obtained by means of a self-mixing interferometry technique (Fig.~\ref{fig_phaseturbulence}d) giving values between 1 and 2 above threshold (Fig.~\ref{fig_phaseturbulence}e) (Fig.~\ref{fig_phaseturbulence}c), which are in very good agreement with values reported in the literature~\cite{Jumpertz2016,Staden2006,Kumazaki_2008}. At a typical pump level of $J/J_{th}=1.1$ for the ring multimode instability, the  extrapolated value of the LEF is $1.9\pm0.5$. The GVD was obtained using the Fourier transform method~\cite{Hofstetter1999} (Fig.~\ref{fig_phaseturbulence}d), which gave a value of $\approx-1800$~fs$^2$/mm (Fig.~\ref{fig_phaseturbulence}e). The $(c_1,c_2)$ coordinates corresponding to our laser parameters are marked in Fig.~\ref{fig_phaseturbulence}a.
The fact that our laser operates in the $(c_1,c_2)$-parameter space close to the Benjamin-Feir line shows that the multimode instability is compatible with the phase turbulence mechanism. We link pump-dependent on- and off-switching of multimode operation to the current-dependent evolution of all the laser parameters, including the GVD and LEF, that can push the system in and out of the instability region.

\begin{figure}[t!]
    \centering
    \includegraphics[width=0.5\textwidth]{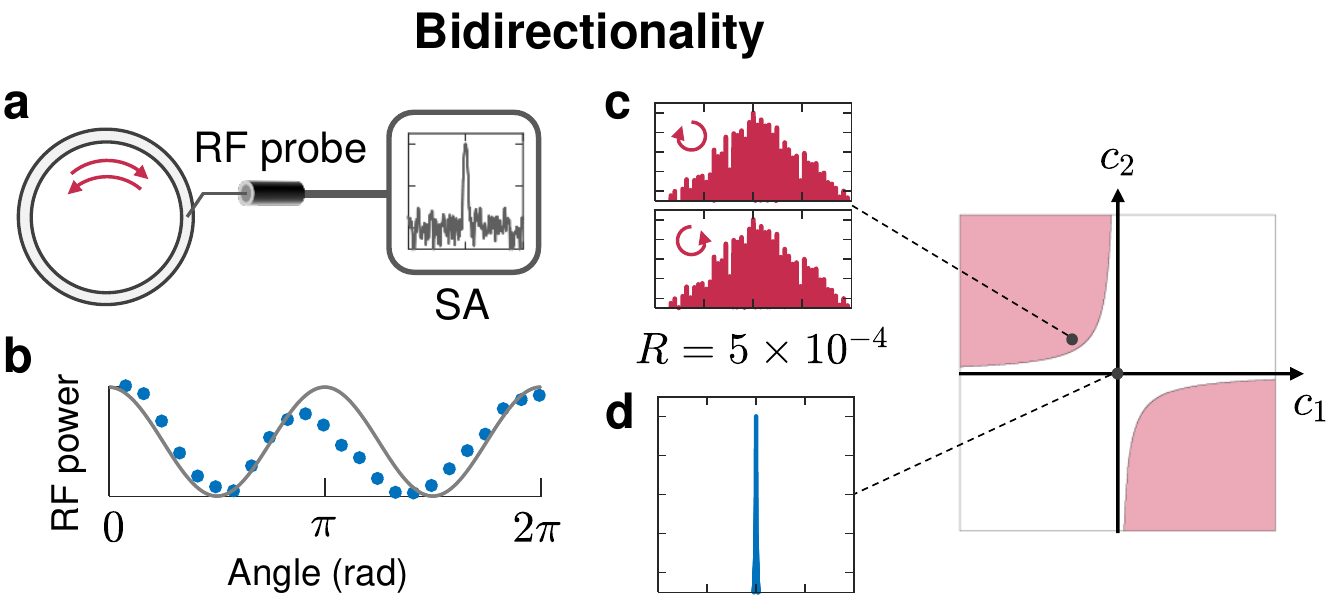}
    \caption{\textbf{a}, Schematic of the technique used to check the occurrence of counter-propagating waves in the ring laser presented in Fig.~\ref{fig_perfectring}c,d. A radiofrequency (RF) probe is scanned along the perimeter of the cavity to measure the beatnote power pattern using a spectrum analyzer (SA). \textbf{b}, Experimental beatnote power pattern (dots) measured at the roundtrip frequency of the laser (27.8~GHz). Also shown is the pattern calculated with an analytical model (line) based on the existence of a bidirectional regime. \textbf{c,d}, Space-time domain simulations of a ring laser with an unintentional defect of small reflectivity, $R=5\times10^{-4}$. In all plots the x-axis spans 100 longitudinal modes of the ring, the y-axis is intensity (10 dB/div). \textbf{c}, When the laser operates in the unstable region of ($c_1$,$c_2$)-parameter space two counter-propagating waves of similar magnitude are formed. \textbf{d}, Instead, when the laser operates in the stable region of parameter space a single mode in a unidirectional regime is observed, indicating that the small defect is not sufficient to trigger multimode operation.}
    \label{fig_bidirectionality}
\end{figure}

Turning our attention to other possible mechanisms intervening in the ring instability, we investigate the directionality of the laser regime. The occurrence of counter-propagating waves in the ring cavity can be verified by means of a radiofrequency (RF) technique that was recently introduced in the context of Fabry-Perot laser frequency combs~\cite{Piccardo:18,PiccardoJSTQE}. This method utilizes an RF probe to measure along the laser cavity the pattern of beat note power that is generated in the frequency comb regime (Fig.~\ref{fig_bidirectionality}a). It has the advantage of being non-invasive, differently from the waveguide outcoupling techniques that are normally used in semiconductor ring lasers to analyze the CW/CCW lasing directions and that may add unwanted reflections~\cite{Gelens2009,Nshii2010}. The beat note power pattern measured in the ring at $f_{rt}$ is shown in Fig.~\ref{fig_bidirectionality}b. In case of a unidirectional mode propagation the beatnote power pattern would remain constant along the waveguide, whereas the presence of nodes indicates that the device operates in a bidirectional regime with optical standing waves in the cavity, as confirmed by an analytical model (Supplementary Material). In order to explain the spatial coupling of counter-propagating waves one must assume that a localized defect is present in the cavity. Unintentional defects may arise from imperfections in the waveguide fabrication but these should have small values of intensity reflectivity $R$. Numerical simulations show that a value as small as $R=5\times10^{-4}$ is sufficient to induce counter-propagating waves when the laser operates in the instability region (Fig.~\ref{fig_bidirectionality}c). At the same time, such small value of $R$ does not induce a multimode regime for points lying in the stable region of parameter space, which continue to exhibit a single mode (Fig.~\ref{fig_bidirectionality}d). We verified that  values of $R$ of at least few percent are needed in the latter case to trigger multimode operation via SHB. We conclude from these results that ring frequency combs are due to the cooperative action of the phase instability, which is responsible in first place for multimode operation, and of an unintentional defect with small reflectivity, which couples the counter-propagating waves in the cavity. The occurrence of multimode operation can be consistently explained by the provided analytical theory, as well as the numerical simulations. We believe that whether a comb or a incoherent multimode state is formed might also depend on further conditions, such as small residual reflections.

\begin{figure}[t!]
    \centering
    \includegraphics[width=0.5\textwidth]{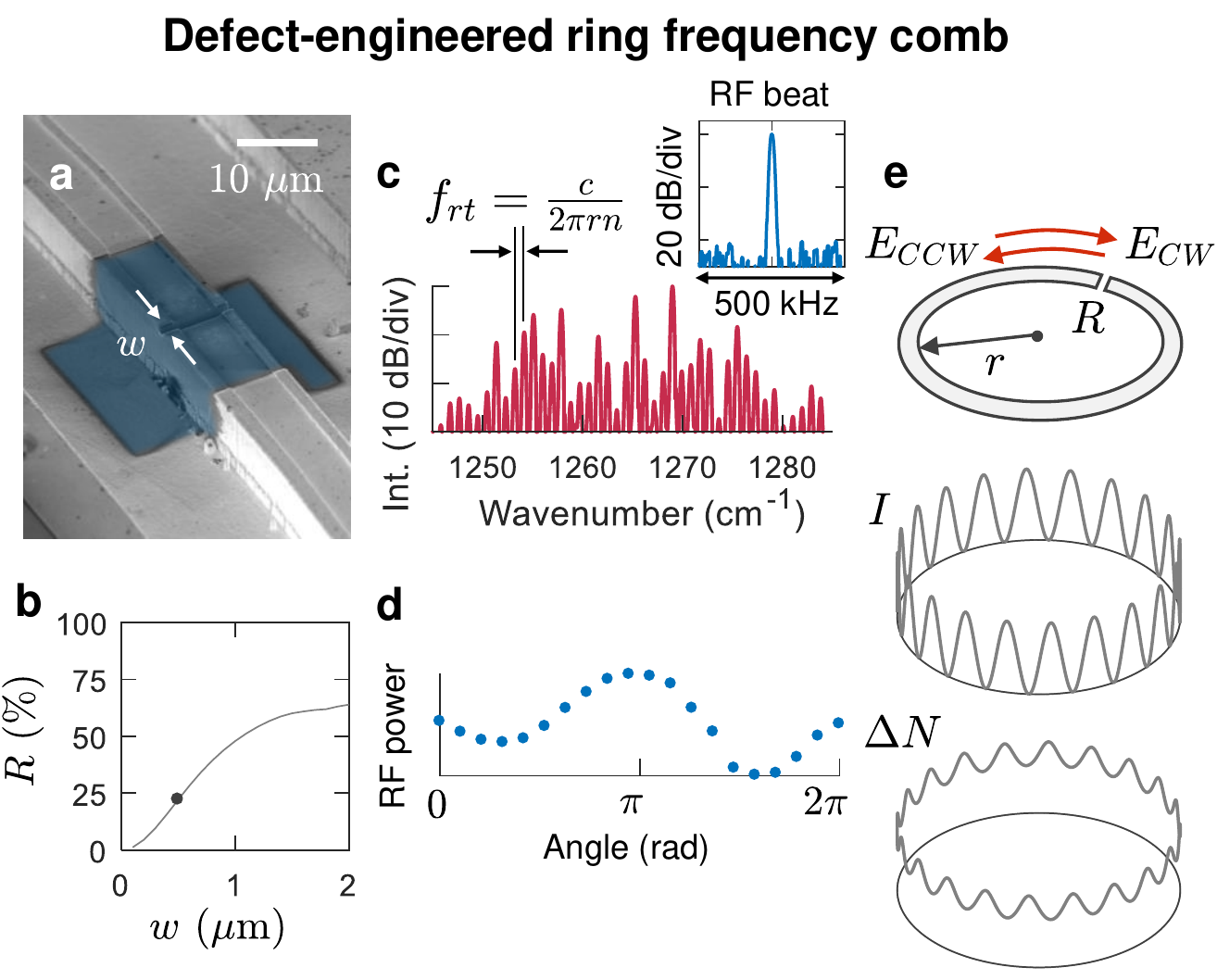}
    \caption{\textbf{a}, Scanning electron microscope image of the defect-engineered ring laser showing the aperture in the metal (blue region in false color) with a 0.5~$\mu$m wide slit fabricated by focused ion beam lithography. \textbf{b}, Reflectivity induced by the slit as a function of its width as calculated from numerical wave simulations. The reflectivity of the studied device is $R\approx 22\%$ (dot). \textbf{c}, Optical spectrum of a defect-engineered ring quantum cascade laser ($r=500$~$\mu$m) fabricated from the same material of the other devices studied in this work. Also shown is the microwave beatnote extracted from the laser (central frequency $\approx$27.8~GHz). \textbf{d}, Experimental beatnote power pattern measured along the perimeter of the ring at the roundtrip frequency. The defect is located where the angle is zero. \textbf{e}, Schematic of the ring laser embedding an engineered defect of reflectivity $R$. The defect induces clockwise and counter-clockwise waves, $E_\mathrm{CW}$ and $E_\mathrm{CCW}$, resulting in an optical standing wave $I$ and a static population grating $\Delta N$.}
    \label{fig_defectengineering}
\end{figure}

To investigate further the role of defects in a ring laser we intentionally embed one in the waveguide by focused ion beam lithography (Fig.~\ref{fig_defectengineering}a). A simple yet effective way of controlling the defect reflectivity is to etch a narrow slit across the waveguide to create an air gap in the active region of the laser. Choosing the slit width in the 0.1-2~$\mu$m interval one can vary the intensity reflectivity $R$ between 1$\%$ and 64$\%$ (Fig.~\ref{fig_defectengineering}b). Our fabricated slit has a width of 0.5~$\mu$m giving $R\approx 22 \%$, which is close to the facet reflectivity of an uncoated Fabry-Perot QCL ($R=29\%$). The defect-engineered laser generates an optical frequency comb producing a narrow beat note at the roundtrip frequency of the laser ($\approx 27.8$~GHz). The optical spectrum exhibits an irregular envelope---the result of complex laser mode competition---similar to that of Fabry-Perot devices, where multimode operation is dominated as well by SHB. The occurrence of counter-propagating waves causing SHB in the device (Fig.~\ref{fig_defectengineering}e) is confirmed by the beatnote power characterization (Fig.~\ref{fig_defectengineering}d), which shows oscillations with a local maximum at the position of the defect, as expected. Defect engineering proves to be a valuable tool for frequency comb generation in ring lasers, as it allows to introduce SHB in a controlled manner in the ring cavity and trigger a multimode instability without inducing significant optical losses as defects can be deeply subwavelength. This technique offers a new degree of freedom in the control of frequency combs that will be investigated further in the future.
At the same time, these results show that ring frequency combs dominated by SHB exhibit different features from those of non-engineered ring lasers. The study of the latter has revealed the phase instability, which is normally masked in presence of SHB. Linking this multimode instability to the CGLE allows one to unravel a possible connection between semiconductor ring lasers and Kerr-driven frequency comb generators. The latter are usually described by the Lugiato-Lefever equations, which are nonlinear Schrödinger equations with well-known soliton solutions derived from the CGLE in the limit of very large $c_1$ and $c_2$ parameters. At the same time we showed that the LEF, which intervenes in both Ginzburg-Landau parameters in the case of a ring laser, effectively contributes as a Kerr nonlinearity. While this is a first step in connecting laser frequency combs together with Kerr combs to the CGLE, we hope that further exploration of the vast ($c_1$,$c_2$) parameter space will strengthen even further the physics shared between these devices.

\bigskip

We acknowledge support from the National Science Foundation under Awards No. CCSS-1807323. Any opinions, findings, conclusions or recommendations expressed in this material are those of the authors and do not necessarily reflect the views of the National Science Foundation. This work was performed in part at the Center for Nanoscale Systems (CNS), a member of the National Nanotechnology Coordinated Infrastructure Network (NNCI), which is supported by the National Science Foundation under NSF Award no. 1541959. B.S. was supported by the Austrian Science Fund (FWF) within the project NanoPlas. We gratefully acknowledge C. A. Wang, M. K. Connors and D. McNulty for providing quantum cascade laser material. We thank V. Ginis and F. Grillot for useful discussions, G. Strasser for allowing the device fabrication, P. Chevalier for cleaving the devices, and D. Ham group for lending us microwave amplifiers.


\begin{thebibliography}{37}%
\makeatletter
\providecommand \@ifxundefined [1]{%
 \@ifx{#1\undefined}
}%
\providecommand \@ifnum [1]{%
 \ifnum #1\expandafter \@firstoftwo
 \else \expandafter \@secondoftwo
 \fi
}%
\providecommand \@ifx [1]{%
 \ifx #1\expandafter \@firstoftwo
 \else \expandafter \@secondoftwo
 \fi
}%
\providecommand \natexlab [1]{#1}%
\providecommand \enquote  [1]{``#1''}%
\providecommand \bibnamefont  [1]{#1}%
\providecommand \bibfnamefont [1]{#1}%
\providecommand \citenamefont [1]{#1}%
\providecommand \href@noop [0]{\@secondoftwo}%
\providecommand \href [0]{\begingroup \@sanitize@url \@href}%
\providecommand \@href[1]{\@@startlink{#1}\@@href}%
\providecommand \@@href[1]{\endgroup#1\@@endlink}%
\providecommand \@sanitize@url [0]{\catcode `\\12\catcode `\$12\catcode
  `\&12\catcode `\#12\catcode `\^12\catcode `\_12\catcode `\%12\relax}%
\providecommand \@@startlink[1]{}%
\providecommand \@@endlink[0]{}%
\providecommand \url  [0]{\begingroup\@sanitize@url \@url }%
\providecommand \@url [1]{\endgroup\@href {#1}{\urlprefix }}%
\providecommand \urlprefix  [0]{URL }%
\providecommand \Eprint [0]{\href }%
\providecommand \doibase [0]{http://dx.doi.org/}%
\providecommand \selectlanguage [0]{\@gobble}%
\providecommand \bibinfo  [0]{\@secondoftwo}%
\providecommand \bibfield  [0]{\@secondoftwo}%
\providecommand \translation [1]{[#1]}%
\providecommand \BibitemOpen [0]{}%
\providecommand \bibitemStop [0]{}%
\providecommand \bibitemNoStop [0]{.\EOS\space}%
\providecommand \EOS [0]{\spacefactor3000\relax}%
\providecommand \BibitemShut  [1]{\csname bibitem#1\endcsname}%
\let\auto@bib@innerbib\@empty
\bibitem [{\citenamefont {Picqu\'e}\ and\ \citenamefont
  {H{\"{a}}nsch}(2019)}]{Picque2019}%
  \BibitemOpen
  \bibfield  {author} {\bibinfo {author} {\bibfnamefont {N.}~\bibnamefont
  {Picqu\'e}}\ and\ \bibinfo {author} {\bibfnamefont {T.~W.}\ \bibnamefont
  {H{\"{a}}nsch}},\ }\href {https://doi.org/10.1038/s41566-018-0347-5}
  {\bibfield  {journal} {\bibinfo  {journal} {Nature Photonics}\ }\textbf
  {\bibinfo {volume} {13}},\ \bibinfo {pages} {146} (\bibinfo {year}
  {2019})}\BibitemShut {NoStop}%
\bibitem [{\citenamefont {Gaeta}\ \emph {et~al.}(2019)\citenamefont {Gaeta},
  \citenamefont {Lipson},\ and\ \citenamefont {Kippenberg}}]{Gaeta2019}%
  \BibitemOpen
  \bibfield  {author} {\bibinfo {author} {\bibfnamefont {A.~L.}\ \bibnamefont
  {Gaeta}}, \bibinfo {author} {\bibfnamefont {M.}~\bibnamefont {Lipson}}, \
  and\ \bibinfo {author} {\bibfnamefont {T.~J.}\ \bibnamefont {Kippenberg}},\
  }\href {https://doi.org/10.1038/s41566-019-0358-x} {\bibfield  {journal}
  {\bibinfo  {journal} {Nature Photonics}\ }\textbf {\bibinfo {volume} {13}},\
  \bibinfo {pages} {158} (\bibinfo {year} {2019})}\BibitemShut {NoStop}%
\bibitem [{\citenamefont {Kues}\ \emph {et~al.}(2019)\citenamefont {Kues},
  \citenamefont {Reimer}, \citenamefont {Lukens}, \citenamefont {Munro},
  \citenamefont {Weiner}, \citenamefont {Moss},\ and\ \citenamefont
  {Morandotti}}]{Kues2019}%
  \BibitemOpen
  \bibfield  {author} {\bibinfo {author} {\bibfnamefont {M.}~\bibnamefont
  {Kues}}, \bibinfo {author} {\bibfnamefont {C.}~\bibnamefont {Reimer}},
  \bibinfo {author} {\bibfnamefont {J.~M.}\ \bibnamefont {Lukens}}, \bibinfo
  {author} {\bibfnamefont {W.~J.}\ \bibnamefont {Munro}}, \bibinfo {author}
  {\bibfnamefont {A.~M.}\ \bibnamefont {Weiner}}, \bibinfo {author}
  {\bibfnamefont {D.~J.}\ \bibnamefont {Moss}}, \ and\ \bibinfo {author}
  {\bibfnamefont {R.}~\bibnamefont {Morandotti}},\ }\href
  {https://doi.org/10.1038/s41566-019-0363-0} {\bibfield  {journal} {\bibinfo
  {journal} {Nature Photonics}\ }\textbf {\bibinfo {volume} {13}},\ \bibinfo
  {pages} {170} (\bibinfo {year} {2019})}\BibitemShut {NoStop}%
\bibitem [{\citenamefont {H{\"{a}}nsch}(2006)}]{Hansch2006}%
  \BibitemOpen
  \bibfield  {author} {\bibinfo {author} {\bibfnamefont {T.~W.}\ \bibnamefont
  {H{\"{a}}nsch}},\ }\href {\doibase 10.1103/RevModPhys.78.1297} {\bibfield
  {journal} {\bibinfo  {journal} {Rev. Mod. Phys.}\ }\textbf {\bibinfo {volume}
  {78}},\ \bibinfo {pages} {1297} (\bibinfo {year} {2006})}\BibitemShut
  {NoStop}%
\bibitem [{\citenamefont {Kippenberg}\ \emph {et~al.}(2011)\citenamefont
  {Kippenberg}, \citenamefont {Holzwarth},\ and\ \citenamefont
  {Diddams}}]{Kippenberg555}%
  \BibitemOpen
  \bibfield  {author} {\bibinfo {author} {\bibfnamefont {T.~J.}\ \bibnamefont
  {Kippenberg}}, \bibinfo {author} {\bibfnamefont {R.}~\bibnamefont
  {Holzwarth}}, \ and\ \bibinfo {author} {\bibfnamefont {S.~A.}\ \bibnamefont
  {Diddams}},\ }\href {\doibase 10.1126/science.1193968} {\bibfield  {journal}
  {\bibinfo  {journal} {Science}\ }\textbf {\bibinfo {volume} {332}},\ \bibinfo
  {pages} {555} (\bibinfo {year} {2011})}\BibitemShut {NoStop}%
\bibitem [{\citenamefont {Herr}\ \emph {et~al.}(2012)\citenamefont {Herr},
  \citenamefont {Hartinger}, \citenamefont {Riemensberger}, \citenamefont
  {Wang}, \citenamefont {Gavartin}, \citenamefont {Holzwarth}, \citenamefont
  {Gorodetsky},\ and\ \citenamefont {Kippenberg}}]{Herr2012}%
  \BibitemOpen
  \bibfield  {author} {\bibinfo {author} {\bibfnamefont {T.}~\bibnamefont
  {Herr}}, \bibinfo {author} {\bibfnamefont {K.}~\bibnamefont {Hartinger}},
  \bibinfo {author} {\bibfnamefont {J.}~\bibnamefont {Riemensberger}}, \bibinfo
  {author} {\bibfnamefont {C.~Y.}\ \bibnamefont {Wang}}, \bibinfo {author}
  {\bibfnamefont {E.}~\bibnamefont {Gavartin}}, \bibinfo {author}
  {\bibfnamefont {R.}~\bibnamefont {Holzwarth}}, \bibinfo {author}
  {\bibfnamefont {M.~L.}\ \bibnamefont {Gorodetsky}}, \ and\ \bibinfo {author}
  {\bibfnamefont {T.~J.}\ \bibnamefont {Kippenberg}},\ }\href@noop {}
  {\bibfield  {journal} {\bibinfo  {journal} {Nature Photonics}\ }\textbf
  {\bibinfo {volume} {6}},\ \bibinfo {pages} {480} (\bibinfo {year}
  {2012})}\BibitemShut {NoStop}%
\bibitem [{\citenamefont {Kaertner}\ \emph {et~al.}(1995)\citenamefont
  {Kaertner}, \citenamefont {Brovelli}, \citenamefont {Kopf}, \citenamefont
  {Kamp}, \citenamefont {Calasso},\ and\ \citenamefont
  {Keller}}]{Kaertner1995}%
  \BibitemOpen
  \bibfield  {author} {\bibinfo {author} {\bibfnamefont {F.~X.}\ \bibnamefont
  {Kaertner}}, \bibinfo {author} {\bibfnamefont {L.~R.}\ \bibnamefont
  {Brovelli}}, \bibinfo {author} {\bibfnamefont {D.}~\bibnamefont {Kopf}},
  \bibinfo {author} {\bibfnamefont {M.}~\bibnamefont {Kamp}}, \bibinfo {author}
  {\bibfnamefont {I.~G.}\ \bibnamefont {Calasso}}, \ and\ \bibinfo {author}
  {\bibfnamefont {U.}~\bibnamefont {Keller}},\ }\href {\doibase
  10.1117/12.204794} {\bibfield  {journal} {\bibinfo  {journal} {Optical
  Engineering}\ }\textbf {\bibinfo {volume} {34}},\ \bibinfo {pages} {2024}
  (\bibinfo {year} {1995})}\BibitemShut {NoStop}%
\bibitem [{\citenamefont {Agrawal}(1988)}]{Agrawal1988}%
  \BibitemOpen
  \bibfield  {author} {\bibinfo {author} {\bibfnamefont {G.~P.}\ \bibnamefont
  {Agrawal}},\ }\href {\doibase 10.1364/JOSAB.5.000147} {\bibfield  {journal}
  {\bibinfo  {journal} {Journal of the Optical Society of America B}\ }\textbf
  {\bibinfo {volume} {5}},\ \bibinfo {pages} {147} (\bibinfo {year}
  {1988})}\BibitemShut {NoStop}%
\bibitem [{\citenamefont {Faist}\ \emph {et~al.}(2016)\citenamefont {Faist},
  \citenamefont {Villares}, \citenamefont {Scalari}, \citenamefont {Rosch},
  \citenamefont {Bonzon}, \citenamefont {Hugi},\ and\ \citenamefont
  {Beck}}]{Faist2016}%
  \BibitemOpen
  \bibfield  {author} {\bibinfo {author} {\bibfnamefont {J.}~\bibnamefont
  {Faist}}, \bibinfo {author} {\bibfnamefont {G.}~\bibnamefont {Villares}},
  \bibinfo {author} {\bibfnamefont {G.}~\bibnamefont {Scalari}}, \bibinfo
  {author} {\bibfnamefont {M.}~\bibnamefont {Rosch}}, \bibinfo {author}
  {\bibfnamefont {C.}~\bibnamefont {Bonzon}}, \bibinfo {author} {\bibfnamefont
  {A.}~\bibnamefont {Hugi}}, \ and\ \bibinfo {author} {\bibfnamefont
  {M.}~\bibnamefont {Beck}},\ }\href {\doibase 10.1515/nanoph-2016-0015}
  {\bibfield  {journal} {\bibinfo  {journal} {Nanophotonics}\ }\textbf
  {\bibinfo {volume} {5}},\ \bibinfo {pages} {272} (\bibinfo {year}
  {2016})}\BibitemShut {NoStop}%
\bibitem [{\citenamefont {Gelens}\ \emph {et~al.}(2009)\citenamefont {Gelens},
  \citenamefont {Beri}, \citenamefont {Van~der Sande}, \citenamefont {Mezosi},
  \citenamefont {Sorel}, \citenamefont {Danckaert},\ and\ \citenamefont
  {Verschaffelt}}]{Gelens2009}%
  \BibitemOpen
  \bibfield  {author} {\bibinfo {author} {\bibfnamefont {L.}~\bibnamefont
  {Gelens}}, \bibinfo {author} {\bibfnamefont {S.}~\bibnamefont {Beri}},
  \bibinfo {author} {\bibfnamefont {G.}~\bibnamefont {Van~der Sande}}, \bibinfo
  {author} {\bibfnamefont {G.}~\bibnamefont {Mezosi}}, \bibinfo {author}
  {\bibfnamefont {M.}~\bibnamefont {Sorel}}, \bibinfo {author} {\bibfnamefont
  {J.}~\bibnamefont {Danckaert}}, \ and\ \bibinfo {author} {\bibfnamefont
  {G.}~\bibnamefont {Verschaffelt}},\ }\href {\doibase
  10.1103/PhysRevLett.102.193904} {\bibfield  {journal} {\bibinfo  {journal}
  {Phys. Rev. Lett.}\ }\textbf {\bibinfo {volume} {102}},\ \bibinfo {pages}
  {193904} (\bibinfo {year} {2009})}\BibitemShut {NoStop}%
\bibitem [{\citenamefont {Risken}\ and\ \citenamefont
  {Nummedal}(1968)}]{Risken1968}%
  \BibitemOpen
  \bibfield  {author} {\bibinfo {author} {\bibfnamefont {H.}~\bibnamefont
  {Risken}}\ and\ \bibinfo {author} {\bibfnamefont {K.}~\bibnamefont
  {Nummedal}},\ }\href {\doibase 10.1063/1.1655817} {\bibfield  {journal}
  {\bibinfo  {journal} {Journal of Applied Physics}\ }\textbf {\bibinfo
  {volume} {39}},\ \bibinfo {pages} {4662} (\bibinfo {year}
  {1968})}\BibitemShut {NoStop}%
\bibitem [{\citenamefont {Graham}\ and\ \citenamefont
  {Haken}(1968)}]{Graham1968}%
  \BibitemOpen
  \bibfield  {author} {\bibinfo {author} {\bibfnamefont {R.}~\bibnamefont
  {Graham}}\ and\ \bibinfo {author} {\bibfnamefont {H.}~\bibnamefont {Haken}},\
  }\href {\doibase 10.1007/BF01405384} {\bibfield  {journal} {\bibinfo
  {journal} {Zeitschrift f{\"{u}}r Physik}\ }\textbf {\bibinfo {volume}
  {213}},\ \bibinfo {pages} {420} (\bibinfo {year} {1968})}\BibitemShut
  {NoStop}%
\bibitem [{\citenamefont {Aranson}\ and\ \citenamefont
  {Kramer}(2002)}]{Aranson2002}%
  \BibitemOpen
  \bibfield  {author} {\bibinfo {author} {\bibfnamefont {I.~S.}\ \bibnamefont
  {Aranson}}\ and\ \bibinfo {author} {\bibfnamefont {L.}~\bibnamefont
  {Kramer}},\ }\href {\doibase 10.1103/RevModPhys.74.99} {\bibfield  {journal}
  {\bibinfo  {journal} {Rev. Mod. Phys.}\ }\textbf {\bibinfo {volume} {74}},\
  \bibinfo {pages} {99} (\bibinfo {year} {2002})}\BibitemShut {NoStop}%
\bibitem [{\citenamefont {Mujagi\'{c}}\ \emph {et~al.}(2008)\citenamefont
  {Mujagi\'{c}}, \citenamefont {Schartner}, \citenamefont {Hoffmann},
  \citenamefont {Schrenk}, \citenamefont {Semtsiv}, \citenamefont {Wienold},
  \citenamefont {Masselink},\ and\ \citenamefont {Strasser}}]{Mujagic2008}%
  \BibitemOpen
  \bibfield  {author} {\bibinfo {author} {\bibfnamefont {E.}~\bibnamefont
  {Mujagi\'{c}}}, \bibinfo {author} {\bibfnamefont {S.}~\bibnamefont
  {Schartner}}, \bibinfo {author} {\bibfnamefont {L.~K.}\ \bibnamefont
  {Hoffmann}}, \bibinfo {author} {\bibfnamefont {W.}~\bibnamefont {Schrenk}},
  \bibinfo {author} {\bibfnamefont {M.~P.}\ \bibnamefont {Semtsiv}}, \bibinfo
  {author} {\bibfnamefont {M.}~\bibnamefont {Wienold}}, \bibinfo {author}
  {\bibfnamefont {W.~T.}\ \bibnamefont {Masselink}}, \ and\ \bibinfo {author}
  {\bibfnamefont {G.}~\bibnamefont {Strasser}},\ }\href@noop {} {\bibfield
  {journal} {\bibinfo  {journal} {Applied Physics Letters}\ }\textbf {\bibinfo
  {volume} {93}},\ \bibinfo {pages} {011108} (\bibinfo {year}
  {2008})}\BibitemShut {NoStop}%
\bibitem [{\citenamefont {Hugi}\ \emph {et~al.}(2012)\citenamefont {Hugi},
  \citenamefont {Villares}, \citenamefont {Blaser}, \citenamefont {Liu},\ and\
  \citenamefont {Faist}}]{Hugi2012}%
  \BibitemOpen
  \bibfield  {author} {\bibinfo {author} {\bibfnamefont {A.}~\bibnamefont
  {Hugi}}, \bibinfo {author} {\bibfnamefont {G.}~\bibnamefont {Villares}},
  \bibinfo {author} {\bibfnamefont {S.}~\bibnamefont {Blaser}}, \bibinfo
  {author} {\bibfnamefont {H.~C.}\ \bibnamefont {Liu}}, \ and\ \bibinfo
  {author} {\bibfnamefont {J.}~\bibnamefont {Faist}},\ }\href {\doibase
  10.1038/nature11620} {\bibfield  {journal} {\bibinfo  {journal} {Nature}\
  }\textbf {\bibinfo {volume} {492}},\ \bibinfo {pages} {229} (\bibinfo {year}
  {2012})}\BibitemShut {NoStop}%
\bibitem [{\citenamefont {Villares}\ \emph {et~al.}(2014)\citenamefont
  {Villares}, \citenamefont {Hugi}, \citenamefont {Blaser},\ and\ \citenamefont
  {Faist}}]{Villares2014}%
  \BibitemOpen
  \bibfield  {author} {\bibinfo {author} {\bibfnamefont {G.}~\bibnamefont
  {Villares}}, \bibinfo {author} {\bibfnamefont {A.}~\bibnamefont {Hugi}},
  \bibinfo {author} {\bibfnamefont {S.}~\bibnamefont {Blaser}}, \ and\ \bibinfo
  {author} {\bibfnamefont {J.}~\bibnamefont {Faist}},\ }\href {\doibase
  10.1038/ncomms6192} {\bibfield  {journal} {\bibinfo  {journal} {Nat.
  Commun.}\ }\textbf {\bibinfo {volume} {5}},\ \bibinfo {pages} {5192}
  (\bibinfo {year} {2014})}\BibitemShut {NoStop}%
\bibitem [{\citenamefont {Consolino}\ \emph {et~al.}(2019)\citenamefont
  {Consolino}, \citenamefont {Nafa}, \citenamefont {Cappelli}, \citenamefont
  {Garrasi}, \citenamefont {Mezzapesa}, \citenamefont {Li}, \citenamefont
  {Davies}, \citenamefont {Linfield}, \citenamefont {Vitiello}, \citenamefont
  {De~Natale},\ and\ \citenamefont {Bartalini}}]{consolino2019fully}%
  \BibitemOpen
  \bibfield  {author} {\bibinfo {author} {\bibfnamefont {L.}~\bibnamefont
  {Consolino}}, \bibinfo {author} {\bibfnamefont {M.}~\bibnamefont {Nafa}},
  \bibinfo {author} {\bibfnamefont {F.}~\bibnamefont {Cappelli}}, \bibinfo
  {author} {\bibfnamefont {K.}~\bibnamefont {Garrasi}}, \bibinfo {author}
  {\bibfnamefont {F.~P.}\ \bibnamefont {Mezzapesa}}, \bibinfo {author}
  {\bibfnamefont {L.}~\bibnamefont {Li}}, \bibinfo {author} {\bibfnamefont
  {A.~G.}\ \bibnamefont {Davies}}, \bibinfo {author} {\bibfnamefont {E.~H.}\
  \bibnamefont {Linfield}}, \bibinfo {author} {\bibfnamefont {M.~S.}\
  \bibnamefont {Vitiello}}, \bibinfo {author} {\bibfnamefont {P.}~\bibnamefont
  {De~Natale}}, \ and\ \bibinfo {author} {\bibfnamefont {S.}~\bibnamefont
  {Bartalini}},\ }\href@noop {} {\bibfield  {journal} {\bibinfo  {journal}
  {arXiv:1902.01604}\ } (\bibinfo {year} {2019})}\BibitemShut {NoStop}%
\bibitem [{\citenamefont {Piccardo}\ \emph {et~al.}(2019)\citenamefont
  {Piccardo}, \citenamefont {Tamagnone}, \citenamefont {Schwarz}, \citenamefont
  {Chevalier}, \citenamefont {Rubin}, \citenamefont {Wang}, \citenamefont
  {Wang}, \citenamefont {Connors}, \citenamefont {McNulty}, \citenamefont
  {Belyanin},\ and\ \citenamefont {Capasso}}]{PiccardoLRT}%
  \BibitemOpen
  \bibfield  {author} {\bibinfo {author} {\bibfnamefont {M.}~\bibnamefont
  {Piccardo}}, \bibinfo {author} {\bibfnamefont {M.}~\bibnamefont {Tamagnone}},
  \bibinfo {author} {\bibfnamefont {B.}~\bibnamefont {Schwarz}}, \bibinfo
  {author} {\bibfnamefont {P.}~\bibnamefont {Chevalier}}, \bibinfo {author}
  {\bibfnamefont {N.~A.}\ \bibnamefont {Rubin}}, \bibinfo {author}
  {\bibfnamefont {Y.}~\bibnamefont {Wang}}, \bibinfo {author} {\bibfnamefont
  {C.~A.}\ \bibnamefont {Wang}}, \bibinfo {author} {\bibfnamefont {M.~K.}\
  \bibnamefont {Connors}}, \bibinfo {author} {\bibfnamefont {D.}~\bibnamefont
  {McNulty}}, \bibinfo {author} {\bibfnamefont {A.}~\bibnamefont {Belyanin}}, \
  and\ \bibinfo {author} {\bibfnamefont {F.}~\bibnamefont {Capasso}},\
  }\href@noop {} {\bibfield  {journal} {\bibinfo  {journal} {Proceedings of the
  National Academy of Sciences}\ }\textbf {\bibinfo {volume} {116}},\ \bibinfo
  {pages} {9181} (\bibinfo {year} {2019})}\BibitemShut {NoStop}%
\bibitem [{\citenamefont {Szedlak}\ \emph {et~al.}(2018)\citenamefont
  {Szedlak}, \citenamefont {Hisch}, \citenamefont {Schwarz}, \citenamefont
  {Holzbauer}, \citenamefont {MacFarland}, \citenamefont {Zederbauer},
  \citenamefont {Detz}, \citenamefont {Andrews}, \citenamefont {Schrenk},
  \citenamefont {Rotter},\ and\ \citenamefont {Strasser}}]{Szedlak2018}%
  \BibitemOpen
  \bibfield  {author} {\bibinfo {author} {\bibfnamefont {R.}~\bibnamefont
  {Szedlak}}, \bibinfo {author} {\bibfnamefont {T.}~\bibnamefont {Hisch}},
  \bibinfo {author} {\bibfnamefont {B.}~\bibnamefont {Schwarz}}, \bibinfo
  {author} {\bibfnamefont {M.}~\bibnamefont {Holzbauer}}, \bibinfo {author}
  {\bibfnamefont {D.}~\bibnamefont {MacFarland}}, \bibinfo {author}
  {\bibfnamefont {T.}~\bibnamefont {Zederbauer}}, \bibinfo {author}
  {\bibfnamefont {H.}~\bibnamefont {Detz}}, \bibinfo {author} {\bibfnamefont
  {A.~M.}\ \bibnamefont {Andrews}}, \bibinfo {author} {\bibfnamefont
  {W.}~\bibnamefont {Schrenk}}, \bibinfo {author} {\bibfnamefont
  {S.}~\bibnamefont {Rotter}}, \ and\ \bibinfo {author} {\bibfnamefont
  {G.}~\bibnamefont {Strasser}},\ }\href
  {https://doi.org/10.1038/s41598-018-26267-x} {\bibfield  {journal} {\bibinfo
  {journal} {Scientific Reports}\ }\textbf {\bibinfo {volume} {8}},\ \bibinfo
  {pages} {7998} (\bibinfo {year} {2018})}\BibitemShut {NoStop}%
\bibitem [{\citenamefont {Malara}\ \emph {et~al.}(2013)\citenamefont {Malara},
  \citenamefont {Blanchard}, \citenamefont {Mansuripur}, \citenamefont
  {Wojcik}, \citenamefont {Belyanin}, \citenamefont {Fujita}, \citenamefont
  {Edamura}, \citenamefont {Furuta}, \citenamefont {Yamanishi}, \citenamefont
  {de~Natale},\ and\ \citenamefont {Capasso}}]{Malara2013}%
  \BibitemOpen
  \bibfield  {author} {\bibinfo {author} {\bibfnamefont {P.}~\bibnamefont
  {Malara}}, \bibinfo {author} {\bibfnamefont {R.}~\bibnamefont {Blanchard}},
  \bibinfo {author} {\bibfnamefont {T.~S.}\ \bibnamefont {Mansuripur}},
  \bibinfo {author} {\bibfnamefont {A.~K.}\ \bibnamefont {Wojcik}}, \bibinfo
  {author} {\bibfnamefont {A.}~\bibnamefont {Belyanin}}, \bibinfo {author}
  {\bibfnamefont {K.}~\bibnamefont {Fujita}}, \bibinfo {author} {\bibfnamefont
  {T.}~\bibnamefont {Edamura}}, \bibinfo {author} {\bibfnamefont
  {S.}~\bibnamefont {Furuta}}, \bibinfo {author} {\bibfnamefont
  {M.}~\bibnamefont {Yamanishi}}, \bibinfo {author} {\bibfnamefont
  {P.}~\bibnamefont {de~Natale}}, \ and\ \bibinfo {author} {\bibfnamefont
  {F.}~\bibnamefont {Capasso}},\ }\href@noop {} {\bibfield  {journal} {\bibinfo
   {journal} {Applied Physics Letters}\ }\textbf {\bibinfo {volume} {102}},\
  \bibinfo {pages} {141105} (\bibinfo {year} {2013})}\BibitemShut {NoStop}%
\bibitem [{\citenamefont {Wojcik}\ \emph {et~al.}(2013)\citenamefont {Wojcik},
  \citenamefont {Malara}, \citenamefont {Blanchard}, \citenamefont
  {Mansuripur}, \citenamefont {Capasso},\ and\ \citenamefont
  {Belyanin}}]{Wojcik2013}%
  \BibitemOpen
  \bibfield  {author} {\bibinfo {author} {\bibfnamefont {A.~K.}\ \bibnamefont
  {Wojcik}}, \bibinfo {author} {\bibfnamefont {P.}~\bibnamefont {Malara}},
  \bibinfo {author} {\bibfnamefont {R.}~\bibnamefont {Blanchard}}, \bibinfo
  {author} {\bibfnamefont {T.~S.}\ \bibnamefont {Mansuripur}}, \bibinfo
  {author} {\bibfnamefont {F.}~\bibnamefont {Capasso}}, \ and\ \bibinfo
  {author} {\bibfnamefont {A.}~\bibnamefont {Belyanin}},\ }\href {\doibase
  10.1063/1.4838275} {\bibfield  {journal} {\bibinfo  {journal} {Applied
  Physics Letters}\ }\textbf {\bibinfo {volume} {103}},\ \bibinfo {pages}
  {231102} (\bibinfo {year} {2013})}\BibitemShut {NoStop}%
\bibitem [{\citenamefont {Revin}\ \emph {et~al.}(2016)\citenamefont {Revin},
  \citenamefont {Hemingway}, \citenamefont {Wang}, \citenamefont {Cockburn},\
  and\ \citenamefont {Belyanin}}]{Revin2016}%
  \BibitemOpen
  \bibfield  {author} {\bibinfo {author} {\bibfnamefont {D.~G.}\ \bibnamefont
  {Revin}}, \bibinfo {author} {\bibfnamefont {M.}~\bibnamefont {Hemingway}},
  \bibinfo {author} {\bibfnamefont {Y.}~\bibnamefont {Wang}}, \bibinfo {author}
  {\bibfnamefont {J.~W.}\ \bibnamefont {Cockburn}}, \ and\ \bibinfo {author}
  {\bibfnamefont {A.}~\bibnamefont {Belyanin}},\ }\href
  {https://doi.org/10.1038/ncomms11440} {\bibfield  {journal} {\bibinfo
  {journal} {Nature Communications}\ }\textbf {\bibinfo {volume} {7}},\
  \bibinfo {pages} {11440} (\bibinfo {year} {2016})}\BibitemShut {NoStop}%
\bibitem [{\citenamefont {Faist}\ \emph {et~al.}(1996)\citenamefont {Faist},
  \citenamefont {Gmachl}, \citenamefont {Striccoli}, \citenamefont {Sirtori},
  \citenamefont {Capasso}, \citenamefont {Sivco},\ and\ \citenamefont
  {Cho}}]{Faist1996}%
  \BibitemOpen
  \bibfield  {author} {\bibinfo {author} {\bibfnamefont {J.}~\bibnamefont
  {Faist}}, \bibinfo {author} {\bibfnamefont {C.}~\bibnamefont {Gmachl}},
  \bibinfo {author} {\bibfnamefont {M.}~\bibnamefont {Striccoli}}, \bibinfo
  {author} {\bibfnamefont {C.}~\bibnamefont {Sirtori}}, \bibinfo {author}
  {\bibfnamefont {F.}~\bibnamefont {Capasso}}, \bibinfo {author} {\bibfnamefont
  {D.~L.}\ \bibnamefont {Sivco}}, \ and\ \bibinfo {author} {\bibfnamefont
  {A.~Y.}\ \bibnamefont {Cho}},\ }\href@noop {} {\bibfield  {journal} {\bibinfo
   {journal} {Applied Physics Letters}\ }\textbf {\bibinfo {volume} {69}},\
  \bibinfo {pages} {2456} (\bibinfo {year} {1996})}\BibitemShut {NoStop}%
\bibitem [{\citenamefont {Gmachl}\ \emph {et~al.}(1998)\citenamefont {Gmachl},
  \citenamefont {Capasso}, \citenamefont {Narimanov}, \citenamefont
  {N{\"o}ckel}, \citenamefont {Stone}, \citenamefont {Faist}, \citenamefont
  {Sivco},\ and\ \citenamefont {Cho}}]{Gmachl1556}%
  \BibitemOpen
  \bibfield  {author} {\bibinfo {author} {\bibfnamefont {C.}~\bibnamefont
  {Gmachl}}, \bibinfo {author} {\bibfnamefont {F.}~\bibnamefont {Capasso}},
  \bibinfo {author} {\bibfnamefont {E.~E.}\ \bibnamefont {Narimanov}}, \bibinfo
  {author} {\bibfnamefont {J.~U.}\ \bibnamefont {N{\"o}ckel}}, \bibinfo
  {author} {\bibfnamefont {A.~D.}\ \bibnamefont {Stone}}, \bibinfo {author}
  {\bibfnamefont {J.}~\bibnamefont {Faist}}, \bibinfo {author} {\bibfnamefont
  {D.~L.}\ \bibnamefont {Sivco}}, \ and\ \bibinfo {author} {\bibfnamefont
  {A.~Y.}\ \bibnamefont {Cho}},\ }\href {\doibase
  10.1126/science.280.5369.1556} {\bibfield  {journal} {\bibinfo  {journal}
  {Science}\ }\textbf {\bibinfo {volume} {280}},\ \bibinfo {pages} {1556}
  (\bibinfo {year} {1998})}\BibitemShut {NoStop}%
\bibitem [{\citenamefont {Wang}\ \emph {et~al.}(2010)\citenamefont {Wang},
  \citenamefont {Yan}, \citenamefont {Yu}, \citenamefont {Unterhinninghofen},
  \citenamefont {Wiersig}, \citenamefont {Pfl{\"u}gl}, \citenamefont {Diehl},
  \citenamefont {Edamura}, \citenamefont {Yamanishi}, \citenamefont {Kan},\
  and\ \citenamefont {Capasso}}]{Wang22407}%
  \BibitemOpen
  \bibfield  {author} {\bibinfo {author} {\bibfnamefont {Q.~J.}\ \bibnamefont
  {Wang}}, \bibinfo {author} {\bibfnamefont {C.}~\bibnamefont {Yan}}, \bibinfo
  {author} {\bibfnamefont {N.}~\bibnamefont {Yu}}, \bibinfo {author}
  {\bibfnamefont {J.}~\bibnamefont {Unterhinninghofen}}, \bibinfo {author}
  {\bibfnamefont {J.}~\bibnamefont {Wiersig}}, \bibinfo {author} {\bibfnamefont
  {C.}~\bibnamefont {Pfl{\"u}gl}}, \bibinfo {author} {\bibfnamefont
  {L.}~\bibnamefont {Diehl}}, \bibinfo {author} {\bibfnamefont
  {T.}~\bibnamefont {Edamura}}, \bibinfo {author} {\bibfnamefont
  {M.}~\bibnamefont {Yamanishi}}, \bibinfo {author} {\bibfnamefont
  {H.}~\bibnamefont {Kan}}, \ and\ \bibinfo {author} {\bibfnamefont
  {F.}~\bibnamefont {Capasso}},\ }\href {\doibase 10.1073/pnas.1015386107}
  {\bibfield  {journal} {\bibinfo  {journal} {Proceedings of the National
  Academy of Sciences}\ }\textbf {\bibinfo {volume} {107}},\ \bibinfo {pages}
  {22407} (\bibinfo {year} {2010})}\BibitemShut {NoStop}%
\bibitem [{\citenamefont {{Wang}}\ \emph {et~al.}(2017)\citenamefont {{Wang}},
  \citenamefont {{Schwarz}}, \citenamefont {{Siriani}}, \citenamefont
  {{Missaggia}}, \citenamefont {{Connors}}, \citenamefont {{Mansuripur}},
  \citenamefont {{Calawa}}, \citenamefont {{McNulty}}, \citenamefont
  {{Nickerson}}, \citenamefont {{Donnelly}}, \citenamefont {{Creedon}},\ and\
  \citenamefont {{Capasso}}}]{WangQCLmaterial2017}%
  \BibitemOpen
  \bibfield  {author} {\bibinfo {author} {\bibfnamefont {C.~A.}\ \bibnamefont
  {{Wang}}}, \bibinfo {author} {\bibfnamefont {B.}~\bibnamefont {{Schwarz}}},
  \bibinfo {author} {\bibfnamefont {D.~F.}\ \bibnamefont {{Siriani}}}, \bibinfo
  {author} {\bibfnamefont {L.~J.}\ \bibnamefont {{Missaggia}}}, \bibinfo
  {author} {\bibfnamefont {M.~K.}\ \bibnamefont {{Connors}}}, \bibinfo {author}
  {\bibfnamefont {T.~S.}\ \bibnamefont {{Mansuripur}}}, \bibinfo {author}
  {\bibfnamefont {D.~R.}\ \bibnamefont {{Calawa}}}, \bibinfo {author}
  {\bibfnamefont {D.}~\bibnamefont {{McNulty}}}, \bibinfo {author}
  {\bibfnamefont {M.}~\bibnamefont {{Nickerson}}}, \bibinfo {author}
  {\bibfnamefont {J.~P.}\ \bibnamefont {{Donnelly}}}, \bibinfo {author}
  {\bibfnamefont {K.}~\bibnamefont {{Creedon}}}, \ and\ \bibinfo {author}
  {\bibfnamefont {F.}~\bibnamefont {{Capasso}}},\ }\href {\doibase
  10.1109/JSTQE.2017.2677899} {\bibfield  {journal} {\bibinfo  {journal} {IEEE
  Journal of Selected Topics in Quantum Electronics}\ }\textbf {\bibinfo
  {volume} {23}},\ \bibinfo {pages} {1} (\bibinfo {year} {2017})}\BibitemShut
  {NoStop}%
\bibitem [{\citenamefont {Gil}\ and\ \citenamefont {Lippi}(2014)}]{Gil2014}%
  \BibitemOpen
  \bibfield  {author} {\bibinfo {author} {\bibfnamefont {L.}~\bibnamefont
  {Gil}}\ and\ \bibinfo {author} {\bibfnamefont {G.~L.}\ \bibnamefont
  {Lippi}},\ }\href@noop {} {\bibfield  {journal} {\bibinfo  {journal} {Phys.
  Rev. Lett.}\ }\textbf {\bibinfo {volume} {113}},\ \bibinfo {pages} {213902}
  (\bibinfo {year} {2014})}\BibitemShut {NoStop}%
\bibitem [{\citenamefont {Opa\v{c}ak}\ and\ \citenamefont
  {Schwarz}(2019)}]{opacak2019theory}%
  \BibitemOpen
  \bibfield  {author} {\bibinfo {author} {\bibfnamefont {N.}~\bibnamefont
  {Opa\v{c}ak}}\ and\ \bibinfo {author} {\bibfnamefont {B.}~\bibnamefont
  {Schwarz}},\ }\href@noop {} {\enquote {\bibinfo {title} {{Theory of frequency
  modulated combs in lasers with spatial hole burning, dispersion and Kerr}},}\
  } (\bibinfo {year} {2019}),\ \Eprint {http://arxiv.org/abs/arXiv:1905.13635}
  {arXiv:1905.13635} \BibitemShut {NoStop}%
\bibitem [{\citenamefont {Chate}(1994)}]{Chate1994}%
  \BibitemOpen
  \bibfield  {author} {\bibinfo {author} {\bibfnamefont {H.}~\bibnamefont
  {Chate}},\ }\href@noop {} {\bibfield  {journal} {\bibinfo  {journal}
  {Nonlinearity}\ }\textbf {\bibinfo {volume} {7}},\ \bibinfo {pages} {185}
  (\bibinfo {year} {1994})}\BibitemShut {NoStop}%
\bibitem [{\citenamefont {Columbo}\ \emph {et~al.}(2018)\citenamefont
  {Columbo}, \citenamefont {Barbieri}, \citenamefont {Sirtori},\ and\
  \citenamefont {Brambilla}}]{Columbo2018}%
  \BibitemOpen
  \bibfield  {author} {\bibinfo {author} {\bibfnamefont {L.~L.}\ \bibnamefont
  {Columbo}}, \bibinfo {author} {\bibfnamefont {S.}~\bibnamefont {Barbieri}},
  \bibinfo {author} {\bibfnamefont {C.}~\bibnamefont {Sirtori}}, \ and\
  \bibinfo {author} {\bibfnamefont {M.}~\bibnamefont {Brambilla}},\ }\href@noop
  {} {\bibfield  {journal} {\bibinfo  {journal} {Opt. Express}\ }\textbf
  {\bibinfo {volume} {26}},\ \bibinfo {pages} {2829} (\bibinfo {year}
  {2018})}\BibitemShut {NoStop}%
\bibitem [{\citenamefont {Jumpertz}\ \emph {et~al.}(2016)\citenamefont
  {Jumpertz}, \citenamefont {Michel}, \citenamefont {Pawlus}, \citenamefont
  {Elsässer}, \citenamefont {Schires}, \citenamefont {Carras},\ and\
  \citenamefont {Grillot}}]{Jumpertz2016}%
  \BibitemOpen
  \bibfield  {author} {\bibinfo {author} {\bibfnamefont {L.}~\bibnamefont
  {Jumpertz}}, \bibinfo {author} {\bibfnamefont {F.}~\bibnamefont {Michel}},
  \bibinfo {author} {\bibfnamefont {R.}~\bibnamefont {Pawlus}}, \bibinfo
  {author} {\bibfnamefont {W.}~\bibnamefont {Elsässer}}, \bibinfo {author}
  {\bibfnamefont {K.}~\bibnamefont {Schires}}, \bibinfo {author} {\bibfnamefont
  {M.}~\bibnamefont {Carras}}, \ and\ \bibinfo {author} {\bibfnamefont
  {F.}~\bibnamefont {Grillot}},\ }\href@noop {} {\bibfield  {journal} {\bibinfo
   {journal} {AIP Advances}\ }\textbf {\bibinfo {volume} {6}},\ \bibinfo
  {pages} {015212} (\bibinfo {year} {2016})}\BibitemShut {NoStop}%
\bibitem [{\citenamefont {von Staden}\ \emph {et~al.}(2006)\citenamefont {von
  Staden}, \citenamefont {Gensty}, \citenamefont {Elsäßer}, \citenamefont
  {Giuliani},\ and\ \citenamefont {Mann}}]{Staden2006}%
  \BibitemOpen
  \bibfield  {author} {\bibinfo {author} {\bibfnamefont {J.}~\bibnamefont {von
  Staden}}, \bibinfo {author} {\bibfnamefont {T.}~\bibnamefont {Gensty}},
  \bibinfo {author} {\bibfnamefont {W.}~\bibnamefont {Elsäßer}}, \bibinfo
  {author} {\bibfnamefont {G.}~\bibnamefont {Giuliani}}, \ and\ \bibinfo
  {author} {\bibfnamefont {C.}~\bibnamefont {Mann}},\ }\href@noop {} {\bibfield
   {journal} {\bibinfo  {journal} {Opt. Lett.}\ }\textbf {\bibinfo {volume}
  {31}},\ \bibinfo {pages} {2574} (\bibinfo {year} {2006})}\BibitemShut
  {NoStop}%
\bibitem [{\citenamefont {Kumazaki}\ \emph {et~al.}(2008)\citenamefont
  {Kumazaki}, \citenamefont {Takagi}, \citenamefont {Ishihara}, \citenamefont
  {Kasahara}, \citenamefont {Sugiyama}, \citenamefont {Akikusa},\ and\
  \citenamefont {Edamura}}]{Kumazaki_2008}%
  \BibitemOpen
  \bibfield  {author} {\bibinfo {author} {\bibfnamefont {N.}~\bibnamefont
  {Kumazaki}}, \bibinfo {author} {\bibfnamefont {Y.}~\bibnamefont {Takagi}},
  \bibinfo {author} {\bibfnamefont {M.}~\bibnamefont {Ishihara}}, \bibinfo
  {author} {\bibfnamefont {K.}~\bibnamefont {Kasahara}}, \bibinfo {author}
  {\bibfnamefont {A.}~\bibnamefont {Sugiyama}}, \bibinfo {author}
  {\bibfnamefont {N.}~\bibnamefont {Akikusa}}, \ and\ \bibinfo {author}
  {\bibfnamefont {T.}~\bibnamefont {Edamura}},\ }\href {\doibase
  10.1143/jjap.47.6320} {\bibfield  {journal} {\bibinfo  {journal} {Japanese
  Journal of Applied Physics}\ }\textbf {\bibinfo {volume} {47}},\ \bibinfo
  {pages} {6320} (\bibinfo {year} {2008})}\BibitemShut {NoStop}%
\bibitem [{\citenamefont {Hofstetter}\ and\ \citenamefont
  {Faist}(1999)}]{Hofstetter1999}%
  \BibitemOpen
  \bibfield  {author} {\bibinfo {author} {\bibfnamefont {D.}~\bibnamefont
  {Hofstetter}}\ and\ \bibinfo {author} {\bibfnamefont {J.}~\bibnamefont
  {Faist}},\ }\href@noop {} {\bibfield  {journal} {\bibinfo  {journal} {IEEE
  Photonics Technology Letters}\ }\textbf {\bibinfo {volume} {11}},\ \bibinfo
  {pages} {1372} (\bibinfo {year} {1999})}\BibitemShut {NoStop}%
\bibitem [{\citenamefont {Piccardo}\ \emph {et~al.}(2018)\citenamefont
  {Piccardo}, \citenamefont {Kazakov}, \citenamefont {Rubin}, \citenamefont
  {Chevalier}, \citenamefont {Wang}, \citenamefont {Xie}, \citenamefont
  {Lascola}, \citenamefont {Belyanin},\ and\ \citenamefont
  {Capasso}}]{Piccardo:18}%
  \BibitemOpen
  \bibfield  {author} {\bibinfo {author} {\bibfnamefont {M.}~\bibnamefont
  {Piccardo}}, \bibinfo {author} {\bibfnamefont {D.}~\bibnamefont {Kazakov}},
  \bibinfo {author} {\bibfnamefont {N.~A.}\ \bibnamefont {Rubin}}, \bibinfo
  {author} {\bibfnamefont {P.}~\bibnamefont {Chevalier}}, \bibinfo {author}
  {\bibfnamefont {Y.}~\bibnamefont {Wang}}, \bibinfo {author} {\bibfnamefont
  {F.}~\bibnamefont {Xie}}, \bibinfo {author} {\bibfnamefont {K.}~\bibnamefont
  {Lascola}}, \bibinfo {author} {\bibfnamefont {A.}~\bibnamefont {Belyanin}}, \
  and\ \bibinfo {author} {\bibfnamefont {F.}~\bibnamefont {Capasso}},\ }\href
  {\doibase 10.1364/OPTICA.5.000475} {\bibfield  {journal} {\bibinfo  {journal}
  {Optica}\ }\textbf {\bibinfo {volume} {5}},\ \bibinfo {pages} {475} (\bibinfo
  {year} {2018})}\BibitemShut {NoStop}%
\bibitem [{\citenamefont {{Piccardo}}\ \emph {et~al.}(2019)\citenamefont
  {{Piccardo}}, \citenamefont {{Kazakov}}, \citenamefont {{Schwarz}},
  \citenamefont {{Chevalier}}, \citenamefont {{Amirzhan}}, \citenamefont
  {{Hillbrand}}, \citenamefont {{AlMutairi}}, \citenamefont {{Wang}},
  \citenamefont {{Xie}}, \citenamefont {{Lascola}}, \citenamefont {{Becker}},
  \citenamefont {{Hildebrandt}}, \citenamefont {{Weih}}, \citenamefont
  {{Belyanin}},\ and\ \citenamefont {{Capasso}}}]{PiccardoJSTQE}%
  \BibitemOpen
  \bibfield  {author} {\bibinfo {author} {\bibfnamefont {M.}~\bibnamefont
  {{Piccardo}}}, \bibinfo {author} {\bibfnamefont {D.}~\bibnamefont
  {{Kazakov}}}, \bibinfo {author} {\bibfnamefont {B.}~\bibnamefont
  {{Schwarz}}}, \bibinfo {author} {\bibfnamefont {P.}~\bibnamefont
  {{Chevalier}}}, \bibinfo {author} {\bibfnamefont {A.}~\bibnamefont
  {{Amirzhan}}}, \bibinfo {author} {\bibfnamefont {J.}~\bibnamefont
  {{Hillbrand}}}, \bibinfo {author} {\bibfnamefont {S.~Z.}\ \bibnamefont
  {{AlMutairi}}}, \bibinfo {author} {\bibfnamefont {Y.}~\bibnamefont {{Wang}}},
  \bibinfo {author} {\bibfnamefont {F.}~\bibnamefont {{Xie}}}, \bibinfo
  {author} {\bibfnamefont {K.}~\bibnamefont {{Lascola}}}, \bibinfo {author}
  {\bibfnamefont {S.}~\bibnamefont {{Becker}}}, \bibinfo {author}
  {\bibfnamefont {L.}~\bibnamefont {{Hildebrandt}}}, \bibinfo {author}
  {\bibfnamefont {R.}~\bibnamefont {{Weih}}}, \bibinfo {author} {\bibfnamefont
  {A.}~\bibnamefont {{Belyanin}}}, \ and\ \bibinfo {author} {\bibfnamefont
  {F.}~\bibnamefont {{Capasso}}},\ }\href {\doibase 10.1109/JSTQE.2019.2908553}
  {\bibfield  {journal} {\bibinfo  {journal} {IEEE Journal of Selected Topics
  in Quantum Electronics}\ }\textbf {\bibinfo {volume} {25}},\ \bibinfo {pages}
  {1} (\bibinfo {year} {2019})}\BibitemShut {NoStop}%
\bibitem [{\citenamefont {Nshii}\ \emph {et~al.}(2010)\citenamefont {Nshii},
  \citenamefont {Ironside}, \citenamefont {Sorel}, \citenamefont {Slight},
  \citenamefont {Zhang}, \citenamefont {Revin},\ and\ \citenamefont
  {Cockburn}}]{Nshii2010}%
  \BibitemOpen
  \bibfield  {author} {\bibinfo {author} {\bibfnamefont {C.~C.}\ \bibnamefont
  {Nshii}}, \bibinfo {author} {\bibfnamefont {C.~N.}\ \bibnamefont {Ironside}},
  \bibinfo {author} {\bibfnamefont {M.}~\bibnamefont {Sorel}}, \bibinfo
  {author} {\bibfnamefont {T.~J.}\ \bibnamefont {Slight}}, \bibinfo {author}
  {\bibfnamefont {S.~Y.}\ \bibnamefont {Zhang}}, \bibinfo {author}
  {\bibfnamefont {D.~G.}\ \bibnamefont {Revin}}, \ and\ \bibinfo {author}
  {\bibfnamefont {J.~W.}\ \bibnamefont {Cockburn}},\ }\href {\doibase
  10.1063/1.3524200} {\bibfield  {journal} {\bibinfo  {journal} {Applied
  Physics Letters}\ }\textbf {\bibinfo {volume} {97}},\ \bibinfo {pages}
  {231107} (\bibinfo {year} {2010})}\BibitemShut {NoStop}%
\end{thebibliography}

%

\end{document}


\title{Supplementary Material to:\\
\medskip
Semiconductor ring laser frequency combs induced by phase turbulence}
	
\author{Marco Piccardo$^\dagger$}
\affiliation{Harvard John A. Paulson School of Engineering and Applied Sciences, Harvard University, Cambridge, MA 02138, USA}

\author{Benedikt Schwarz$^\dagger$}
\affiliation{Harvard John A. Paulson School of Engineering and Applied Sciences, Harvard University, Cambridge, MA 02138, USA}
\affiliation{Institute of Solid State Electronics, TU Wien, 1040 Vienna, Austria}

\author{Dmitry Kazakov}
\affiliation{Harvard John A. Paulson School of Engineering and Applied Sciences, Harvard University, Cambridge, MA 02138, USA}

\author{Maximilian Beiser}
\affiliation{Institute of Solid State Electronics, TU Wien, 1040 Vienna, Austria}

\author{Nikola Opacak}
\affiliation{Institute of Solid State Electronics, TU Wien, 1040 Vienna, Austria}

\author{Yongrui Wang}
\affiliation{Department of Physics and Astronomy, Texas A\&M University, College Station, TX 77843, USA}

\author{Shantanu Jha}
\affiliation{Harvard John A. Paulson School of Engineering and Applied Sciences, Harvard University, Cambridge, MA 02138, USA}
\affiliation{Physics Department, Yale University, New Haven, CT 06511, USA}

\author{Michele Tamagnone}
\affiliation{Harvard John A. Paulson School of Engineering and Applied Sciences, Harvard University, Cambridge, MA 02138, USA}

\author{Wei Ting Chen}
\affiliation{Harvard John A. Paulson School of Engineering and Applied Sciences, Harvard University, Cambridge, MA 02138, USA}

\author{Alexander Y. Zhu}
\affiliation{Harvard John A. Paulson School of Engineering and Applied Sciences, Harvard University, Cambridge, MA 02138, USA}

\author{Lorenzo L. Columbo}
\affiliation{Dipartimento di Elettronica e Telecomunicazioni, Politecnico di Torino, Corso Duca degli Abruzzi 24, 10129 Torino, Italy}

\author{Alexey Belyanin}
\affiliation{Department of Physics and Astronomy, Texas A\&M University, College Station, TX 77843, USA}

\author{Federico Capasso}
\email[]{capasso@seas.harvard.edu}
\affiliation{Harvard John A. Paulson School of Engineering and Applied Sciences, Harvard University, Cambridge, MA 02138, USA}

\collaboration{$^\dagger$These authors contributed equally to this work.}

\maketitle

\section{\large{M\lowercase{aterials and methods}}}

\textbf{Quantum cascade lasers.} The lasers emit around 8~$\mu$m and have a structure consisting of GaInAs/AlInAs layers on an InP substrate. The waveguide width is 10~$\mu$m. They are operated under constant electrical injection with a low-noise current driver (Wavelength Electronics QCL LAB 1500), and their temperature is stabilized at 16$^\circ$C using a low-thermal-drift temperature controller (Wavelength Electronics TC5). The threshold current density of a symmetric ring (no defect, 600~$\mu$m radius) was observed to be as low as 1.1~kA/cm$^2$, while that of the defect-engineered ring (500~$\mu$m radius) was 1.3~kA/cm$^2$. The small increase in threshold current density in the defect-engineered ring is attributed to the losses induced by the defect. For comparison the threshold current density of a Fabry-Perot device fabricated from the same material, with the same waveguide width and cleaved to have approximately the same length of the symmetric ring ($L=$3.7~mm) was 1.4~kA/cm$^2$. Only a small amount of power ($\lesssim1$~mW) is scattered out from the ring waveguide, minimizing the perturbations due to outcoupling on the intrinsic states of the lasers. The laser spectral output is measured using a Fourier transform infrared spectrometer and a sensitive photodetector (HgCdTe detector cooled at 77~K). Beat notes produced during frequency comb operation are electrically extracted from the laser chip using a radiofrequency probe connected to a spectrum analyzer. Since the group velocity dispersion (GVD) and linewidth enhancement factor (LEF) characterizations rely on techniques~\cite{Hofstetter1999,Staden2006,Jumpertz2016} for which the output power from a ring would be insufficient, we measure these quantities using Fabry-Perot devices fabricated from the same material and having the same waveguide width of the rings (10~$\mu$m). More details on the LEF characterization are given in the corresponding section of the Supplementary Material.

\textbf{Defect engineering.} 
The width of the defect cut by FIB across the ring waveguide is 500~nm. The reflectivity of the defect is calculated using the frequency domain electromagnetic wave model (\textit{emw} module) of COMSOL using $n=3.19$ as the effective refractive index of the QCL waveguide. We confirmed the values of defect reflectivity obtained from COMSOL simulations with a calculation using the transfer matrix formalism for a dielectric-air-dielectric interface. The reflectivity peak corresponds to a defect width of approximately quarter wavelength in air ($\approx 2$~$\mu$m) and its value ($64\%$) is dictated by the air-dielectric index contrast. If needed, higher values of reflectivity could be achieved by defining a distributed Bragg reflector (DBR) section in the waveguide using FIB milling.

\textbf{Radiofrequency gratings.}  For the measurement of the radiofrequency (or dynamic) gratings we use a coaxial RF probe (Quater A-20338) mounted on an XYZ micrometer positioning stage and placed in contact with the top electrode of the rings. The scanning probe is manually positioned along the perimeter of the ring laser cavity~\cite{Piccardo:18}. The signal detected from the probe at every position is amplified with a RF amplifier (CTT ALN 300-8023, bandwidth 18-26.5~GHz, 22~dB gain) and recorded with a spectrum analyzer (Agilent E4448A). The bandwidth of the probe (DC-18~GHz) is smaller than the typical beat note frequency of the ring lasers (23-27~GHz), however the extracted RF signal is still largely sufficient to characterize the beat note power distribution along the ring cavity.

\section{\large{D\lowercase{erivation of the complex \uppercase{G}inzburg-\uppercase{L}andau equation form of the \uppercase{QCL}s master equation}}}
The complex Ginzburg-Landau equation (CGLE) is one of the most known nonlinear equations in physics. It describes the dynamics of spatially extended system of oscillators near Hopf bifurcation in a qualitative, and often in a quantitative way. The range of the physics fields where it is used includes superconductivity, superfluidity, Bose-Einstein condensates, quantum field theory and laser theory~\cite{aranson2002ginzburg}. In this section, we will show that CGLE can be used to describe the evolution of the electric field inside the laser cavity. We will furthermore reduce the parameter space that goes into the laser model to just two dimensions. In certain regions of this parameter space, the laser can exhibit a multimode instability known as the phase turbulence regime in the CGLE usual notation. Additionally, we will derive an analytical condition that defines the onset of this instability. 

We start from Maxwell-Bloch equations, which we will be able to convert to the CGLE form. Maxwell-Bloch equations~\cite{Shimoda1984Introduction} are commonly used to analyze the evolution of the laser system dynamics. This method, obtained from coupled density matrix formalism and Maxwell's equations, is fully capable of quantitatively describing the coherence of the system. However, due to its complexity, it does not provide an intuitive understanding of the underlying physical processes. For this reason, we will turn to a recent theoretical work~\cite{opacak2019theory}, where Maxwell-Bloch equations were reduced to a single master equation, derived for fast gain media, such as QCLs. We will use it as a starting point to obtain the CGLE and include the so-called linewidth enhancement factor (LEF), which describes the carrier-induced index changes. The master equation reads:

\begin{equation}
	\label{eq_master}
	\begin{aligned}
    \frac{n}{c}& \partial_t E_{\pm}  \pm \partial_x E_{\pm} = i\frac{k''}{2}\partial^2_t E_{\pm} +i\beta (|E_+|^2+|E_-|^2)E_{\pm} \\
    &+\frac{g_0}{2 \Big(  1+ \frac{|E_+|^2+|E_-|^2}{E^2_{sat}} \Big) } \frac{1+i\alpha}{1+i\xi} \Bigg\{ E_\pm - \frac{T_2}{1+i\xi}\partial_t E_\pm \\ &+\Big(\frac{T_2}{1+i\xi}\Big)^2\partial^2_t E_\pm - \frac{T_g}{T_1 E^2_{sat}} \bigg[  E_\pm|E_\mp|^2 \\
    &-\partial_t E_\pm|E_\mp|^2 \Big( T_g + \frac{T_2}{1+i\xi} \Big) - E_\pm E_\mp \partial_t E^*_\mp \Big( T_g+ \\
    &+\frac{T_2}{1+i\xi}\Big) - E_\pm E^*_\mp \partial_t E_\mp\frac{T_2}{1+i\xi} \bigg]  \Bigg\} - \frac{\alpha_w}{2}E_\pm
    \end{aligned}
\end{equation}

Here $E_\pm$ are the complex amplitudes of the two counter-propagating waves in the laser cavity, $n$ is the refractive index, $c$ the speed of light, $T_1$ the carrier non-radiative lifetime, $T_2$ the dephasing time, $T_g=(T^{-1}_1+4k^2D)^{-1}$ is the grating lifetime, where $k$ and $D$ stand for the wavenumber and diffusion coefficient. We have introduced furthermore $k''$ the group velocity dispersion coefficient (GVD), $\beta$ the Kerr non-linearity and $\alpha$ the linewidth enhancement factor with its auxiliary functions $\xi=(\sqrt{\alpha^2+1}-1)/\alpha$ and $a_1=(1+\alpha\xi)/(1+\xi^2)$. The linewidth enhancement factor is defined as $\alpha=-\partial\chi_r/\partial N / (\partial\chi_i/\partial N)$, where $\chi_{r,i}$ are the real and imaginary part of the linear susceptibility, and $N$ is the carrier density. The saturation field is $E^2_{sat}=(2\hbar^2)/(\mu^2T_1T_2a_1)$, where $\mu$ is the dipole matrix element. The power loss coefficient is $\alpha_w$ and the unsaturated gain is $g_0=(\Gamma\mu^2\omega_0T_1T_2J)/(\hbar n\varepsilon_0cL)$, with $\Gamma$ the confinement factor, $J$ the pumping current normalized to the electron charge and $L$ the QCL active region period length. To avoid potential inconsistencies, we use $~e^{+i\omega t}$ as the convention for the direction of the time propagation, since Eq.~\ref{eq_master} uses the same. To study the instability onset in a ring cavity, we can analyze the propagation of a unidirectional field. This is justified, since we are considering a cavity without defects, close to threshold. We can switch to a unidirectional field by setting, for example, $E_-$ to zero. Furthermore, we will use just $E$ to refer to $E_+$, for simplicity:

\begin{equation}
	\label{eq_unidir}
	\begin{aligned}
    \frac{n}{c}& \partial_t E  + \partial_x E = 
    \frac{g(P)}{2} \frac{1+i\alpha}{1+i\xi} \bigg[ E - \frac{T_2}{1+i\xi}\partial_t E \\ &+\Big(\frac{T_2}{1+i\xi}\Big)^2\partial^2_t E   \bigg] - \frac{\alpha_w}{2}E +i\frac{k''}{2}\partial^2_t E +i\beta |E|^2E  ,
    \end{aligned}
\end{equation}
where $g(P)=g_0/(1+P/P_{sat})$ is the saturated gain with normalized power $P=|E|^2$ and saturation power $P_{sat}=E^2_{sat}$. 

Before moving on, we introduce modifications to Eq.~\ref{eq_unidir}. We will neglect the term $\frac{T_2}{1+i\xi}\partial_tE$ in square brackets, since its main effect is to modify the propagation speed of the wave and close to the multimode instability, it is not influential. Secondly, one needs to keep in mind that the master equation is derived from the Maxwell-Bloch equations that utilize the rotating-wave approximation~\cite{Shimoda1984Introduction}. Therefore switching the second time derivative $\partial^2_t$ with a second spatial derivative $ \frac{c^2}{n^2}\partial^2_x$ is an excellent approximation. Moreover, the system is switched to a moving frame of reference by applying the coordinate transformations $x\xrightarrow{}x-\frac{c}{n}t$ and $t\xrightarrow{}t$. We now have:

\begin{equation}
	\label{eq_unidir1}
	\begin{aligned}
    \frac{n}{c}& \partial_t E  = 
    \frac{g(P)}{2} \frac{1+i\alpha}{1+i\xi} \bigg[ E    +\Big(\frac{T_2}{1+i\xi}\Big)^2 \frac{c^2}{n^2} \partial^2_x E   \bigg]\\ &- \frac{\alpha_w}{2}E +i\frac{k''}{2}\frac{c^2}{n^2} \partial^2_x E +i\beta |E|^2E  ,
    \end{aligned}
\end{equation}

It is now of use if we write the Taylor expansion of the saturable gain around the stationary power $P_0$ in the form of:

\begin{equation}
	\label{eq_expand_gain}
    g(P) = g_1 - g_2\frac{|E|^2}{P_{sat}}  ,
\end{equation}
where we have introduced the gain coefficients:

\begin{equation}
    \begin{aligned}
	\label{eq_g1_g2}
    &g_1=\frac{1}{a_1}\frac{\alpha_w}{g_{0,\alpha}}(2g_{0,\alpha}-\alpha_w) ,\\
    &g_2=\frac{1}{a_1}\frac{\alpha^2_w}{g_{0,\alpha}},
    \end{aligned}
\end{equation}
with the modified unsaturated gain due to LEF being $g_{0,\alpha}=g_0a_1$.

We will furthermore introduce new functions $a_1,a_2,...a_6$ for simplicity. First,

\begin{equation}
    \begin{aligned}
	\label{eq_a1_a2}
	&\frac{1+i\alpha}{1+i\xi}=a_1+ia_2, \\
    &a_1=\frac{1+\alpha\xi}{1+\xi^2} ,\\
    &a_2=\frac{\alpha-\xi}{1+\xi^2},
    \end{aligned}
\end{equation}
and
\begin{equation}
    \begin{aligned}
	\label{eq_a3_a4}
	&\Big(\frac{1}{1+i\xi}\Big)^2=a_3+ia_4, \\
    &a_3=\frac{1-\xi^2}{(1+\xi^2)^2} ,\\
    &a_4=-\frac{2\xi}{(1+\xi^2)^2} , \\
    &a_5=a_1a_3-a_2a_4 ,\\
    &a_6=a_1a_4+a_2a_3
    \end{aligned}
\end{equation}

We can write then

\begin{equation}
    \begin{aligned}
	\label{eq_a5_a6}
	\Big( \frac{1+i\alpha}{1+i\xi} \Big) \bigg[ 1&+\Big( \frac{1}{1+i\xi} \Big)^2  \bigg] = a_1+ia_2+(a_5+ia_6)
    \end{aligned}
\end{equation}

After inserting Eqs.~\ref{eq_g1_g2} and \ref{eq_a5_a6} in Eq. \ref{eq_unidir1} and after some derivation, it is possible to get the following form of the master equation:

\begin{equation}
	\label{eq_cgle_1}
	\begin{aligned}
    \partial_t E  = (\gamma+i\omega_s)E + (d_R+id_I)\partial^2_x E -(n_R+in_I)|E|^2E ,
    \end{aligned}
\end{equation}
where we introduced following functions:
\begin{equation}
	\label{eq_func1}
	\begin{aligned}
    &\gamma=\frac{g_1a_1-\alpha_w}{2}\frac{c}{n} , \\
    &\omega_s =\frac{g_1a_2}{2}\frac{c}{n} , \\
    &d_R = \frac{g_1a_5}{2}\frac{c^3}{n^3}T^2_2 , \\
    &d_I =  \frac{g_1a_6T^2_2+k''}{2} \frac{c^3}{n^3} , \\
    &n_R = \frac{g_2a_1}{2P_{sat}}\frac{c}{n} , \\
    &n_I = \Big( \frac{g_2a_2}{2P_{sat}}-\beta \Big)\frac{c}{n}
    \end{aligned}
\end{equation}

Here coefficient $\gamma$ determines the gain, $\omega_s$ is the frequency shift due to the gain asymmetry induced by the LEF, $(d_R+id_I)$ gives the complex diffusion coefficient and $(n_R+in_I)$ describes the nonlinearity.

As a final step before obtaining the CGLE, we will introduce the following transformations:

\begin{equation}
	\label{eq_trans}
	\begin{aligned}
    &t \xrightarrow{} \frac{t}{\gamma} ,\\
    &x \xrightarrow{} \Big(\frac{d_R}{\gamma} \Big)^{1/2}x, \\
    & E \xrightarrow{}  \Big(\frac{\gamma}{n_R} \Big)^{1/2}e^{i\omega_st/\gamma}E,
    \end{aligned}
\end{equation}
and define parameters $c_1$ and $c_2$ that reduce the parameter space:

\begin{equation}
	\label{eq_c1_c2}
	\begin{aligned}
    &c_1= \frac{d_I}{d_R} = \frac{g_1a_6T^2_2+k''}{g_1a_5T^2_2} ,\\
    &c_2 = \frac{n_I}{n_R} = \frac{g_2a_2-2P_{sat}\beta}{g_2a_1}
    \end{aligned}
\end{equation}

It is important to note that GVD influences the value of $c_1$, the Kerr non-linearity figures in $c_2$, while the LEF gives a contribution to both $c_1$ and $c_2$ through the terms $a_1,a_2,...,a_6$. This will be easily seen in the next subsection, in the approximation for small LEF values.
After applying Eqs.~\ref{eq_trans} and \ref{eq_c1_c2}, Eq.~\ref{eq_cgle_1} transforms to the conventional form of the CGLE often found in literature:

\begin{equation}
	\label{eq_cgle}
	\begin{aligned}
    \partial_t E  = E + (1+ic_1)\partial^2_x E -(1+ic_2)|E|^2E .
    \end{aligned}
\end{equation}

The purpose of obtaining the CGLE equations starting from the laser master equation (Eq.~\ref{eq_master}) was to find the origin of the low-threshold multimode instability that occurs in QCL ring cavity devices, which could not be explained with standard theory, e.g. multimode instability induced by spatial hole burning which dominates in Fabry-Perot cavities~\cite{opacak2019theory}. One could do so by analyzing the stability of the CGLE by assuming a plane wave solution, corresponding to a laser operating in a single mode regime. By adding a small perturbation to the plane wave, one can locate the regions of stability of the $(c_1,c_2)$ parameter space by determining if the perturbation is decaying over time (see Ref.~\cite{aranson2002ginzburg}). With such analysis, one can obtain the so-called Benjamin-Fair-Newell criterion:
\begin{equation}
	\label{eq_BFN}
	\begin{aligned}
    1+c_1c_2>0 ,
    \end{aligned}
\end{equation}
which, if fulfilled, states that the plane wave solution is stable and the laser remains in the single mode regime. Otherwise, a band of wavenumbers emerges, within which an instability occurs and gives rise to a state that is called the phase turbulence state. This state is characterized by the competition between multiple different wavenumbers. This means that a laser can go past its multimode instability threshold depending only on the values of the parameters such as the GVD, Kerr or the LEF---if the $1+c_1c_2 < 0$ criterion \ref{eq_BFN} is fulfilled. No additional effect (e.g. spatial hole burning) is required.

\textbf{Small LEF approximation.} From the definitions of the parameters $c_1$ and $c_2$ in Eq.~\ref{eq_c1_c2}, the influence of the GVD and Kerr nonlinearity is obvious. However, contribution from the LEF enters through the terms $a_1,a_2,...,a_6$ and this dependence is not trivial. We will here consider the case where the value of the LEF is small, because relations will become significantly simpler.

We consider small values of $\alpha$ so that the approximation $\xi \approx \alpha/2$ is valid to a reasonable extent. We can then approximate $a_1,a_2,...,a_6$ to the first order of $\alpha$ as $a_1 \approx 1$, $a_2 \approx \alpha/2$, $a_3 \approx 1$, $a_4 \approx -\alpha$, $a_5 \approx 1$ and $a_6 \approx - \alpha/2$. Replugging these approximations back to eq.~\ref{eq_c1_c2} yields the following relations:

\begin{equation}
	\label{eq_c1_c2_app}
	\begin{aligned}
    &c_1= \frac{d_I}{d_R} \approx -\frac{\alpha}{2}+\frac{g_0}{\alpha_w(2g_0-\alpha_w)T^2_2}k'' ,\\
    &c_2 = \frac{n_I}{n_R} \approx \frac{\alpha}{2} - \frac{2P_{sat}g_0}{\alpha^2_w}\beta.
    \end{aligned}
\end{equation}
 
 From the previous relation, which is valid for small values of $\alpha$, it is now clear how the GVD influences $c_1$ and the Kerr nonlinearity influences $c_2$, while the LEF influences both of them. The dependencies on these parameters are linear for both $c_1$ and $c_2$.
 
 Furthermore, from the definition of $c_1$ and $c_2$ one additional conclusion can be drawn. The fact that the LEF figures in the expression for $c_2$ indicates that the LEF can contribute as a Kerr-type term to the phase instability. All of this is due to their fast gain dynamics. Namely, in the derivation of the master equation (Eq.~\ref{eq_master}), it is implied that the carrier lifetimes are sufficiently short so that the carrier population can follow the variations of the intracavity field. It is exactly this effect in combination with the gain saturation that results in a finite contribution to the Kerr term from the LEF. If it were otherwise, as in lasers with considerably larger carrier lifetimes, we would not be able to use the master equation. An additional equation for the carrier population would have to be introduced. Because of slower dynamics, the carrier population would not be able anymore to follow the modulations of the intracavity field and would effectively see only its average value. This would result, in a laser with a slow gain medium, in a minuscule contribution from the LEF to the Kerr term. One can conclude that QCLs serve as unique laser sources, as they represent an ideal platform for studying the rich physics behind the frequency comb formation, at the same time bridging the gap between microresonator combs that arise from the Kerr nonlinearity and more standard semiconductor laser frequency combs.

\section{\large{P\lowercase{arameters used in numerical simulations of ring \uppercase{QCL}s}}}

\begin{tabular}{ |p{1.2cm}||p{4.3cm}||p{2.2cm}|  }
 \hline
 Symbol& Description & Value\\
 \hline
 $T_{ul}$ & Upper-lower level lifetime   & $1 \: \mathrm{ps} $ \\
$T_{ug}$ & Upper-ground level lifetime     & $6 \: \mathrm{ps} $ \\
$T_{lg}$ & Lower-ground level lifetime     & $0.08 \: \mathrm{ps} $ \\
$T_{2}$ & Dephasing time  & $60 \: \mathrm{fs} $ \\
$n$ & Refractive index  & $3.3 $ \\
$D$ & Diffusion coefficient  & $46\: \mathrm{cm}^2/s $ \\
$\alpha_{w}$ & Waveguide power losses   & $4 \: \mathrm{cm}^{-1} $ \\
$\mu$ & Dipole matrix element  & $1 \: \mathrm{nm}\times\mathrm{e}  $ \\
$n_{tot}$ & Sheet density  & $6\!\times10^{10} \: \mathrm{cm}^{\!-2}  $ \\
$\Gamma$ & Confinement factor   & $0.5 $ \\
$L$ & Period length   & $580 \: \angstrom $ \\
$L_c$ & Cavity length   & $4 \: \mathrm{mm} $ \\
$\lambda_0$ & Central wavelength   & $8 \: \mathrm{\mu m} $ \\
 \hline
\end{tabular}

\section{\large{A\lowercase{nalytical model of dynamic gratings in a ring laser with a defect}}}

\begin{figure}[b]
    \centering
    \includegraphics[width=0.25\textwidth]{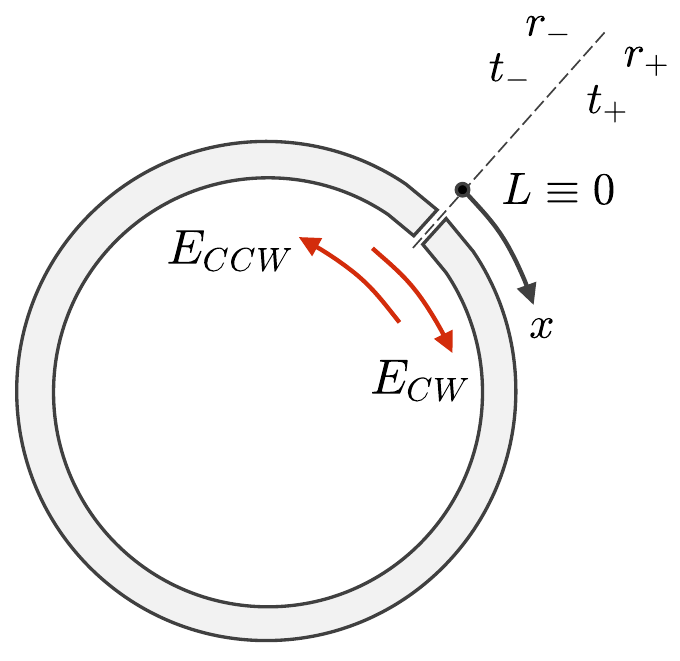}
    \caption{Coordinate system for the analytical model of dynamic gratings in a ring with a defect. The defect is located at the point corresponding to the aperture in the schematic of the ring. The directions of the two counter-propagating waves are shown, as well as the notation for the reflection and transmission coefficients of the defect.}
    \label{fig_analytical_grating_coords}
\end{figure}

\begin{figure*}[t]
    \centering
    \includegraphics[width=0.8\textwidth]{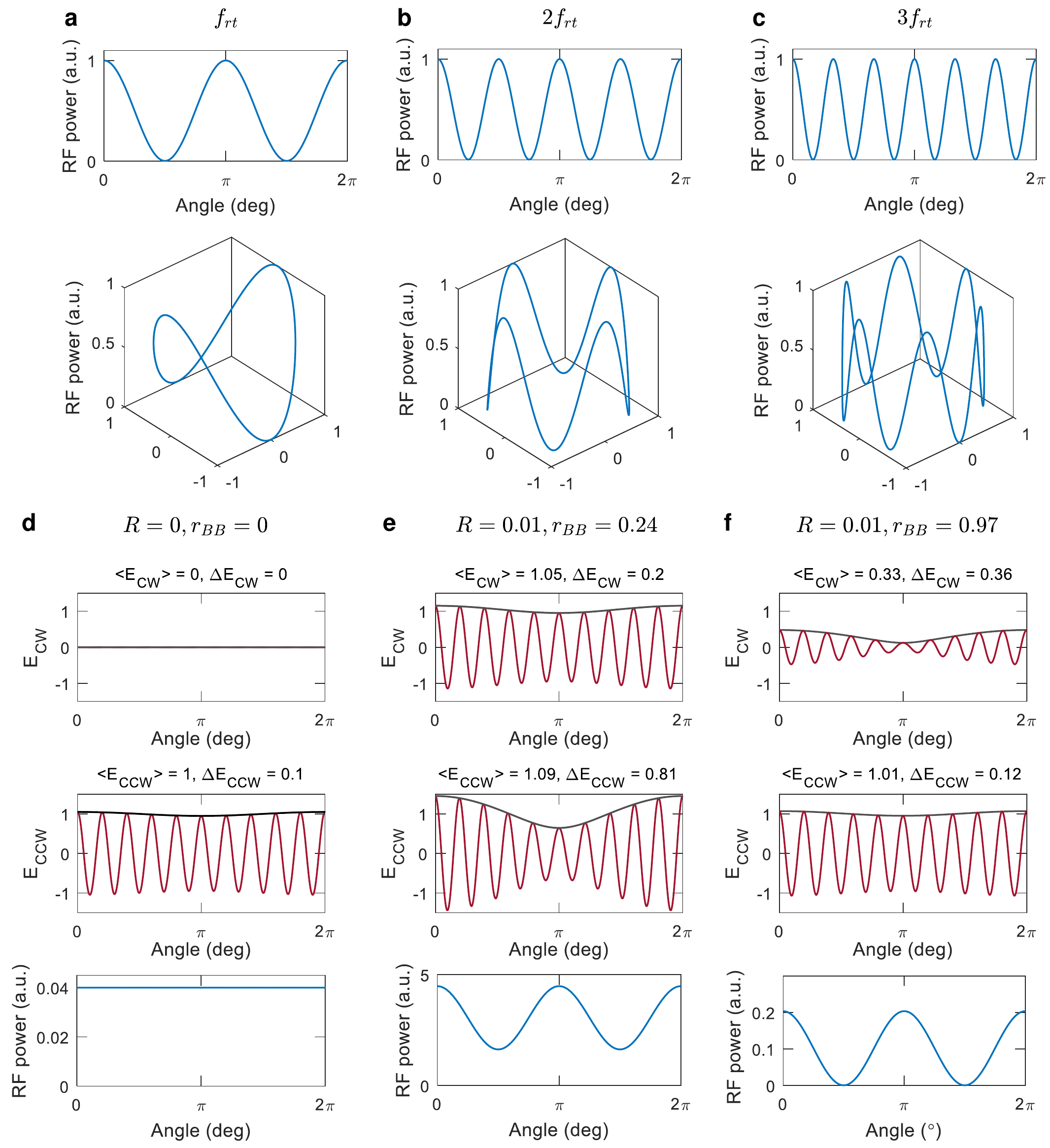}
    \caption{\textbf{a-c}, Beat patterns calculated from the analytical model of a ring with a defect that oscillate at the fundamental, second harmonic and third harmonic of the roundtrip frequency. Patterns are shown both for the unwrapped angular coordinate (top) and as projected onto a 2D ring (bottom). Here it is assumed that the counter-propagating optical beats have the same intensity. \textbf{d-f}, Different beat patterns calculated assuming various beat balance ratios $r_{BB}$, i.e. different relative intensities of the counter-propagating optical beats, as discussed in the text. Also shown are the electric fields of the clockwise and counter-clockwise waves (red curves). The wavenumber is small for visual representation. The black lines correspond to the envelope of the fields from which the mean values $\langle E \rangle$ and modulations amplitudes $\Delta E$ are calculated. The three cases correspond to: \textbf{d}, unidirectional lasing, which gives a uniform beat power across the cavity; \textbf{e}, bidirectional lasing with counter-propagating optical beats not fully balanced, which gives a beat grating with limited fringe visibility; \textbf{f}, bidirectional lasing with fully balanced optical beats, which gives a dynamic grating with strongly suppressed nodes.}
    \label{fig_gratings_orders}
\end{figure*}

In this section we derive an analytical model that allows to predict the shape of the oscillatory gratings occurring in an injection ring laser with a defect operating in a frequency comb regime. The time-dependent population inversion gratings originate from the beating of the optical modes. An analytical model of this phenomenon was already derived in the case of a Fabry-Perot laser in Ref.~\cite{Piccardo:18}. The main difference here is that a ring requires periodic boundary conditions for the field circulating inside the cavity. This results in peculiar differences in terms of the beat patterns among the two types of cavities. Moreover we will show that, differently from the case of a Fabry-Perot cavity, the fringe visibility of the beat pattern of a ring is not uniquely defined for a cavity with given reflective boundaries, but it depends also on the relative magnitude of the counter-propagating waves.

We start by considering two counter-propagating waves, $E_\mathrm{CW}$ and $E_\mathrm{CCW}$, in a ring cavity of perimeter $L$ with a defect of arbitrary reflection and transmission coefficients, $r_{+,-}$ and $t_{+,-}$ (Fig.~\ref{fig_analytical_grating_coords}). The relation among the last four coefficients can be fixed assuming conservation of power at the defect, as it will be discussed later. The two electric fields can be written as

\begin{equation}
	\label{eq_counter+1}
    E_\mathrm{CW}(x,t) = A_\mathrm{CW} e^{\frac{g x}{2}} e^{i(k x - \omega t)} + \textrm{c.c.}
\end{equation}
\begin{equation}
	\label{eq_counter-1}
    E_\mathrm{CCW}(x,t) = A_\mathrm{CCW} e^{\frac{g(L-x)}{2}} e^{i(k x + \omega t)} + \textrm{c.c.}
\end{equation}

where $g$ is the net gain of the medium. Differently from a Fabry-Perot cavity, in a ring the free spectral range is given by $c/(L n_w)$, where $n_{w}$ is the effective group index of the waveguide. Thus the wavevector of the cavity modes can take values given by $k_m = 2\pi m/L$, where $m$ is a positive integer, with the corresponding angular frequency being $\omega_m = c k_m/n_{w}$. The relation among the field amplitudes $A_\mathrm{CW}$ and $A_\mathrm{CCW}$ will be deduced using the boundary conditions

\begin{equation}
	\label{eq_BC1}
    E_\mathrm{CW}(0,t) = E_\mathrm{CCW}(0,t) r_+ + E_\mathrm{CW}(L,t) t_-
\end{equation}
\begin{equation}
	\label{eq_BC2}
    E_\mathrm{CCW}(L,t) = E_\mathrm{CW}(L,t) r_- + E_\mathrm{CCW}(0,t) t_+.
\end{equation}

By plugging Eqs.~\ref{eq_counter+1}, \ref{eq_counter-1} into Eqs.~\ref{eq_BC1}, \ref{eq_BC2} one can rewrite explicitly the boundary conditions in terms of the counter-propagating waves as

\begin{equation}
    \label{eq_BC1_expl}
    \begin{aligned}
    A_\mathrm{CW} \large( e^{-i \omega t} + c.c. \large) & = A_\mathrm{CCW} e^{g L/2} \large( e^{i \omega t} + c.c. \large) r_+ \\
    & + A_\mathrm{CW} e^{g L/2} \large[ e^{i ( kL - \omega t )} + c.c. \large] t_-
    \end{aligned}
\end{equation}

\begin{equation}
    \label{eq_BC2_expl}
    \begin{aligned}
    A_\mathrm{CCW} \large[ e^{i(kL + \omega t)} + c.c. \large] & = A_\mathrm{CW} e^{g L/2} \large[ e^{i (kL - \omega t)} + c.c. \large] r_-\\
    & + A_\mathrm{CCW} e^{g L/2} \large( e^{i \omega t} + c.c. \large) t_+
    \end{aligned}
\end{equation}

The term in brackets require particular attention. Since in a ring wavevectors are such that $k L$ is always a multiple of $2\pi$ one can use the fact that

\begin{equation}
    e^{i(kL+\omega t)} + c.c. =  e^{i(kL - \omega t)} + c.c. =  e^{i\omega t} + c.c.
\end{equation}

to eliminate all the time-dependent terms from Eqs.~\ref{eq_BC1_expl}, \ref{eq_BC2_expl}. This gives the relation between the field amplitudes
\begin{equation}
    \label{eqn_fieldsrelation}
    A_\mathrm{CW} = A_\mathrm{CCW} \frac{e^{gL/2} r_+}{1-e^{gL/2} t_-}
\end{equation}
as well as a transcendental equation that can be solved numerically for a given set of reflection and transmission coefficients to obtain the gain coefficient
\begin{equation}
    1-e^{gL/2} t_+ = \frac{e^{gL/2} r_+}{1-e^{gL/2} t_-} e^{gL/2} r_-.
\end{equation}

Concerning the reflection and transmission coefficients, by assuming conservation of energy at the defect
\begin{equation}
    |E_\mathrm{CW}(L,t)|^2 + |E_\mathrm{CCW}(0,t)|^2 = |E_\mathrm{CW}(0,t)|^2 + |E_\mathrm{CCW}(L,t)|^2
\end{equation}
which implies that the power getting into the defect comes out of the defect, one obtains in combination with Eqs.~\ref{eq_BC1}, \ref{eq_BC2} the following relations
\begin{equation}
    \label{eq_coeff1}
    r_-t_+ + r_+t_- = 0
\end{equation}
\begin{equation}
    \label{eq_coeff2}
    |r_-|^2 = 1 - |t_-|^2
\end{equation}
\begin{equation}
\label{eq_coeff3}
    |r_+|^2 = 1 - |t_+|^2
\end{equation}

By setting any of the coefficients, e.g. $r_+ = \mathrm{cos}(\theta)>0$, the others are fixed by physical considerations. In the limit of small defect reflection coefficient one wants the fraction in Eq.~\ref{eqn_fieldsrelation} to be small, which requires $t_- \approx -1$. (Note that in the assumption of energy conservation at the defect, $g=0$ because there are no mirror losses in a ring.) To satisfy Eqs.~\ref{eq_coeff1}, \ref{eq_coeff2}, \ref{eq_coeff3}, this results in the following set of coefficients: $r_+ = \mathrm{cos}(\theta)$, $r_- = \mathrm{cos}(\theta)$, $t_+ = \mathrm{sin}(\theta)$, $t_-=-\mathrm{sin}(\theta)$. Here $\theta$ is the only unbound parameter of the defect given by $\mathrm{acos}(\sqrt{R})$, where $R$ is the intensity reflectivity.

At this point all the parameters are determined and by setting the amplitude of a wave running in one direction, e.g. $A_\mathrm{CCW}$, one can use Eq.~\ref{eqn_fieldsrelation} to calculate the amplitude of the counter-propagating wave as induced by the defect reflection. Finally a given mode of the cavity with wavevector $k_m$ can be calculated from the sum of Eqs.~\ref{eq_counter+1}, \ref{eq_counter-1} as
\begin{equation}
    E_m(x,t) = E_{\mathrm{CW},m}(x,t) + E_{\mathrm{CCW},m}(x,t).
\end{equation}

The minimum number of independent modes that should be considered to study dynamic gratings in the ring is four, namely the amplitudes of two modes should be set for each propagation direction in order to create two counter-propagating optical beats in the cavity. Then, the mechanism converting the optical beats into a microwave or higher frequency beat note remains the same as the one described in Ref.~\cite{Piccardo:18}. Gratings of different order can be studied depending on the frequency difference among these modes. Fig.~\ref{fig_gratings_orders}a-c shows the power of dynamic gratings oscillating at the roundtrip frequency ($f_{rt}$), and its second and third harmonic, respectively. Here we assume that the counter-propagating optical beats have the same intensity. For a dense frequency comb spaced by one free spectral range (FSR) of the resonator, this is equivalent to considering the beating among first-, second-, and third-order neighbors of the optical modes. In other words, all such gratings can co-exist in a laser but can be discerned based on their characteristic oscillation frequency. The main difference with respect to the dynamic gratings of Fabry-Perot lasers~\cite{Piccardo:18} is that the number of spatial cycles in a ring is always even, for any beat note order. This is due to the periodic boundary conditions of the ring resonator (cf. Fig.~1 of the main text).

Another important difference between dynamic gratings in Fabry-Perot and ring cavities is that the fringe visibility of a beat pattern cannot be solely deduced from the parameters characterizing the reflecting interfaces of the resonator, namely the facets reflectivity in a Fabry-Perot and the defect reflectivity in a ring. In fact, the fringe visibility depends also on the relative intensity of the counter-propagating optical beats, which we call here the beat balance ratio $r_{BB}=I_{\mathrm{CW},B}/I_{\mathrm{CCW},B}$. While in an (uncoated) Fabry-Perot cavity $r_{BB}=1$ because of the symmetry of the resonator that forces counter-propagating waves to be mirror images of each other, in a ring cavity $r_{BB}$ can in principle assume any value. For instance, in the case of a unidirectional regime~\cite{Malara2013}, where there is no defect ($R=0$) and only a wave running in one direction exists, $r_{BB}=0$ and the beat power is uniform along the cavity (Fig.~\ref{fig_gratings_orders}d).For non-zero values of $r_{BB}$ a dynamic grating is formed. A defect must be present ($R\neq0$) in order to lock the phases of the counter-propagating waves. When $r_{BB} \ll 1$ the fringe visibility is limited and the nodes of the pattern have finite power (Fig.~\ref{fig_gratings_orders}e). On the other hand, when $r_{BB}\approx 1$ the visibility is high and the nodes correspond to zeros of beat power (Fig.~\ref{fig_gratings_orders}f).

A subtle point that we want to highlight here is that a pattern with high fringe visibility can be produced even in presence of two counter-propagating waves with very different mean intensity values. This is shown in Fig.~\ref{fig_gratings_orders}f, where the CCW wave has considerably larger intensity than the CW wave but the resulting beat pattern exhibits nodes with strong suppression. This apparent contradiction is explained by the fact that the intensity of the optical beats depends on both the mean value of the field envelope (noted $\langle E_\mathrm{CW} \rangle$ and $\langle E_\mathrm{CCW} \rangle$) and its amplitude modulation (noted $\Delta E_\mathrm{CW}$ and $\Delta E_\mathrm{CCW}$). In particular the beat balance ratio can be written in first approximation as
\begin{equation}
    r_{BB} = \frac{I_{\mathrm{CW},B}}{I_{\mathrm{CCW},B}} \approx \frac{\langle E_\mathrm{CW} \rangle \Delta E_\mathrm{CW}}{\langle E_\mathrm{CCW} \rangle \Delta E_\mathrm{CCW}}
\end{equation}
if one neglects the contribution of higher beat frequencies. To put it in simple terms, the reason why the pattern in Fig.~\ref{fig_gratings_orders}f has high fringe visibility is because the CCW wave is stronger but more weakly modulated than the CW wave. These considerations also indicate an important aspect of the light-matter interaction in the ring laser. In presence of a defect with small reflectivity, the optical coupling between the two counter-propagating waves is small. However the two waves are still coupled through population pulsation in the gain medium and can balance their optical beats. Such type of balanced states are also observed in numerical space-time domain simulations of the QCL based on the laser equations and they hint to an interesting dynamics of the laser that could be further studied in the future.

\section{\large{O\lowercase{n the measurement of dynamic gratings}}}
\label{rf_gratings}

Signal-only RF probe (Quater A-20338) allows to map microwave carrier gratings that arise due to interaction of counter-propagating frequency comb modes with the laser gain medium both in FP and in ring cavities. For both geometries probe is placed on top of the gold contact adjacent to the waveguide and is scanned along it. At each contact point, after amplification (CTT ALN/300-8023), the RF spectrum within the span near the fundamental comb intermodal beat note frequency is acquired and its peak power is recorded. We note that the results of such measurement can be corrupted by spurious reflections of the microwaves propagating on the chip \cite{PiccardoJSTQE}. Therefore, observing some regular structure in the obtained pattern one may falsely assume that the laser cavity supports coupling of the counterpropagating optical modes, while in reality this structure could arise due only to on-chip RF propagation. Whereas for a Fabry-Perot cavity and a defect-engineered ring cavity the consistency of the patterns with the locations of the cavity boundaries suggests that measured patterns reflect the dynamics of the intrinsic microwave carrier gratings, in case of a symmetric ring with seemingly no boundaries extra verification measurements are needed to rule out the possibility of aforementioned artifacts. In absence of perfect circular symmetry any features arising in the measured profiles of the ring laser may be due to chip irregularities, such as, for example, wire bonds that inevitably have to be attached to the chip. We show in this example how wirebond placement can alter the beatnote spatial structure: we first place five wirebonds at the edge of the ring and record the pattern, that shows strong suppression of the the microwave power in the region where wire bonds are attached. We next compare the outcome with the case when the wire bonds are placed in the middle of the ring: the angular pattern is now drastically different and resembles the theoretical prediction for the ring with a defect as pinning point. To ensure that the pattern is due to coupled counter-propagating CW and CCW modes and not to RF standing waves on the chip we also map the radial pattern through the center of the ring: it is symmetric with lobes that decline exponentially towards the middle of the ring, which suggests that RF waves do not propagate atop the chip, but get rapidly attenuated. It is thus reasonable to conclude that even in the ring without any defects with perfect circular symmetry counter-propagating modes are coupled by some arbitrary defect that is seen by the circulating optical mode as discussed in the main text of the manuscript.
\begin{figure}[h]
    \centering
    \includegraphics[width=0.40\textwidth]{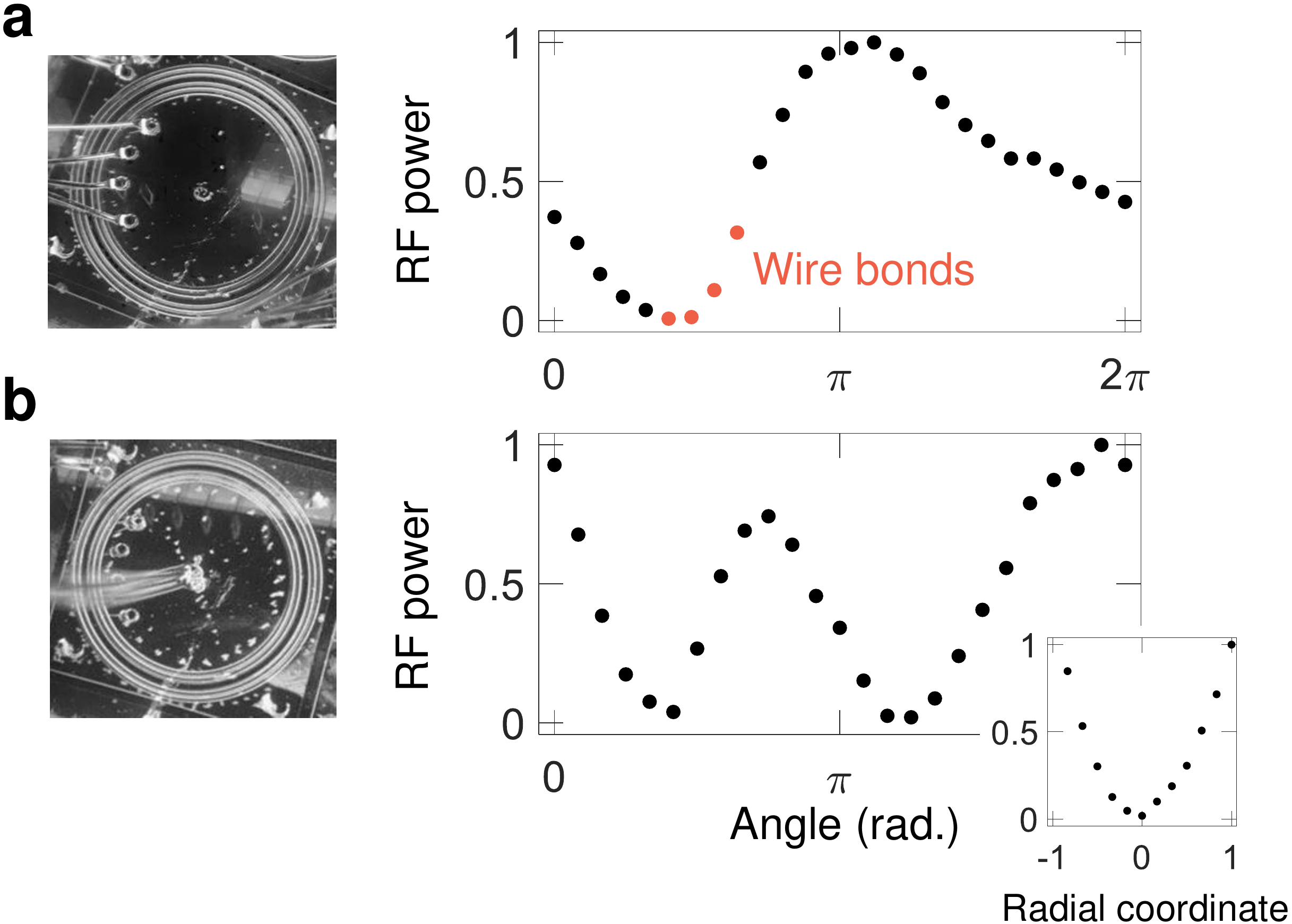}
    \caption{\textbf{a}, Optical microscope image of a symmetric ring QCL with wire bonds attached to the edge of the chip. The RF beatnote pattern is affected by spurious reflections induced by the asymmetric wire bond placement. \textbf{b}, Same laser chip as in \textbf{a} with wire bonds attached to the center of the ring. Intrinsic pattern is not altered thanks to the circular symmetry of the structure. Inset shows the beatnote power as a function of radial coordinate (origin is in the center of the ring) when RF probe is scanned across a diameter of the ring: pattern is symmetric with the beatnote being significantly weaker in the middle of the chip.}
    \label{fig_rf_gratings_wirebonds}
\end{figure}

\section{\large{S\lowercase{elf-interferometry measurements for the linewidth enhancement factor}}}
\label{sec_LEF}

\textbf{Theory.} By self-mixing a Fabry-Perot quantum cascade laser with modulated optical feedback, we are able to observe interferometry effects containing information about its LEF, also known as the $\alpha$-factor of the laser. As has been previously demonstrated from Lang and Kobayashi equations and three-mirror cavity descriptions of optical feedback,
we can describe the the single-mode behaviour of the Fabry-Perot QCL under optical feedback, with the following equations,

\begin{equation}\label{alpha factor equations}
\begin{cases}
    \phi_F(\tau_{ext}) &= \phi_0(\tau_{ext}) - C\sin[\phi_F(\tau_{ext}) + \arctan(\alpha)]\\ \phi_0(\tau_{ext}) &= \omega_0\tau_{ext}\\ \phi_F(\tau_{ext}) &= \omega_F(\tau_{ext})\tau_{ext}\\
    G(\phi_F(\tau_{ext})) &= \cos(\phi_F(\tau_{ext}))\\
    P\left(\phi_F(\tau_{ext})\right) &= P_0\left[1 + mG(\phi_F(\tau_{ext})\right]
\end{cases}
\end{equation}
where $\tau_{ext}$ is the external cavity round trip time, $\alpha$ is the LEF, $C$ is the optical feedback coupling constant, $P_0$, $\phi_0(\tau_{ext})$ and $\omega_0$ are the signal power, phase and angular frequency of the free-running laser, $P(\phi_F(\tau_{ext}))$, $\phi_F(\tau_{ext})$ and $\omega_F(\tau_{ext})$ are the signal power, phase and angular frequency of the laser with optical feedback, and $m$ is a scaling parameter of modulation. 

\autoref{alpha factor equations} can be rewritten as follows,
\begin{equation}\label{phi_F numerical calc}
    \omega_F(\tau_{ext})\tau_{ext} = \omega_0\tau_{ext} - C\sin[\omega_F\tau_{ext} + \arctan(\alpha)]
\end{equation}

From numerical simulation and past work,
we find that Eq.~\ref{phi_F numerical calc} has one solution for $\omega_F(\tau_{ext})$ when $C < 1$ and multiple solutions when $C > 1$, at some $\tau_{ext}$ values. This translates to hysteresis in the modulated power of the self-interfered QCL vs. $\tau_{ext}$. In the experiment, we must then check that our laser is in the weak feedback regime of $C < 1$ to prevent distortion due to hysteresis in the interferogram. 

\begin{figure}[h]
    \centering
    \includegraphics[width=.49\textwidth]{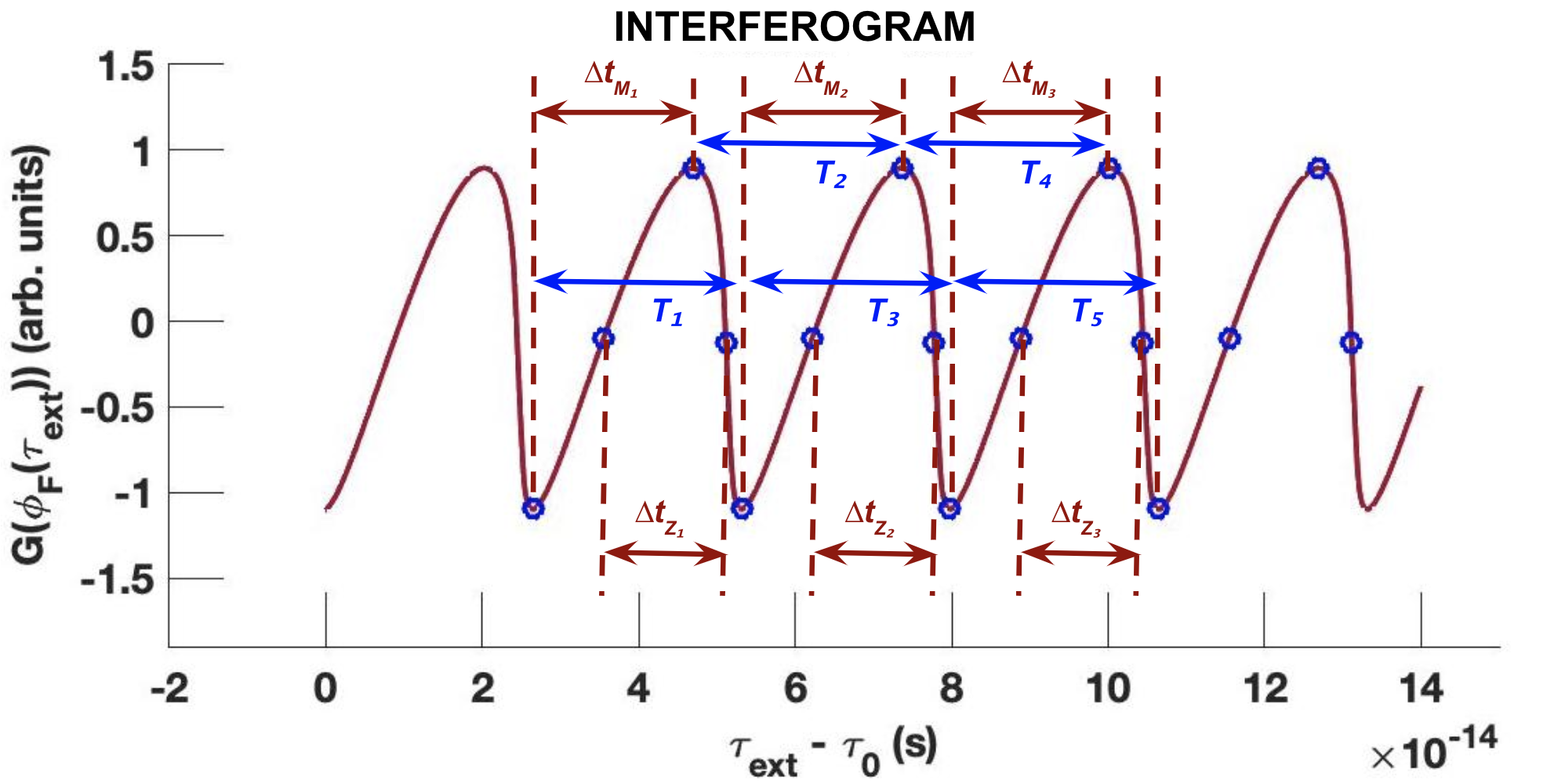}
    \caption{Smulated interferogram fringes produced with $C = 0.9$ and $\alpha = 3$. The asymmetry of these fringes contains information about both $C$ and $\alpha$ that we can extract using $\Delta t_M$ and $\Delta t_Z$.}
    \label{fig_alpha_factor_calculation}
\end{figure}

We show simulated interferogram fringes in Fig.~\ref{fig_alpha_factor_calculation} for $C = 0.9$ and $\alpha = 3$. As studied before in QCLs~\cite{Jumpertz2016,Staden2006}, we can use $T$, $\Delta t_M$ and $\Delta t_Z$, defined in Eq.~\ref{fig_alpha_factor_calculation}, to calculate $\alpha$ and $C$,

\begin{equation}\label{mean alpha}
\begin{cases}
    \overline{\alpha} &= \frac{\overline{\Delta t_M} - \frac{1}{2}\overline{T}}{\overline{\Delta t_Z} - \frac{1}{2}\overline{T}}\\ \overline{C} &= \left|\frac{\overline{\Delta t_M} - \frac{1}{2}\overline{T}}{2\sin(\arctan(\overline{\alpha}))}\frac{2\pi}{\overline{T}}\hspace{.32mm}\right|
\end{cases}
\end{equation}
where a bar denotes the mean-value of the respective variable. Both the sign and magnitude of alpha are found using Eq.~\ref{mean alpha} on data that is plotted versus linearly increasing $\tau_{ext}$, as later shown in numerical simulation.

As an example, we propagate error to $\alpha$ as follows:
\begin{equation}\label{error in alpha}
{\scriptstyle
    \delta \alpha = \sqrt{\left(\pdv{\alpha}{T}\right)^2\left(\delta T\right)^2 + \left(\pdv{\alpha}{\Delta t_M}\right)^2\left(\delta \Delta t_M\right)^2 + \left(\pdv{\alpha}{\Delta t_Z}\right)^2\left(\delta \Delta t_Z\right)^2}
}
\end{equation}
where $\delta T$, $\delta \Delta t_M$, and $\delta \Delta t_Z$ are experimental uncertainties calculated from multiple fringes in the same interferogram, as shown in Fig.~\ref{fig_alpha_factor_calculation}.

\begin{figure}[h]
    \centering
    \includegraphics[width=0.4\textwidth]{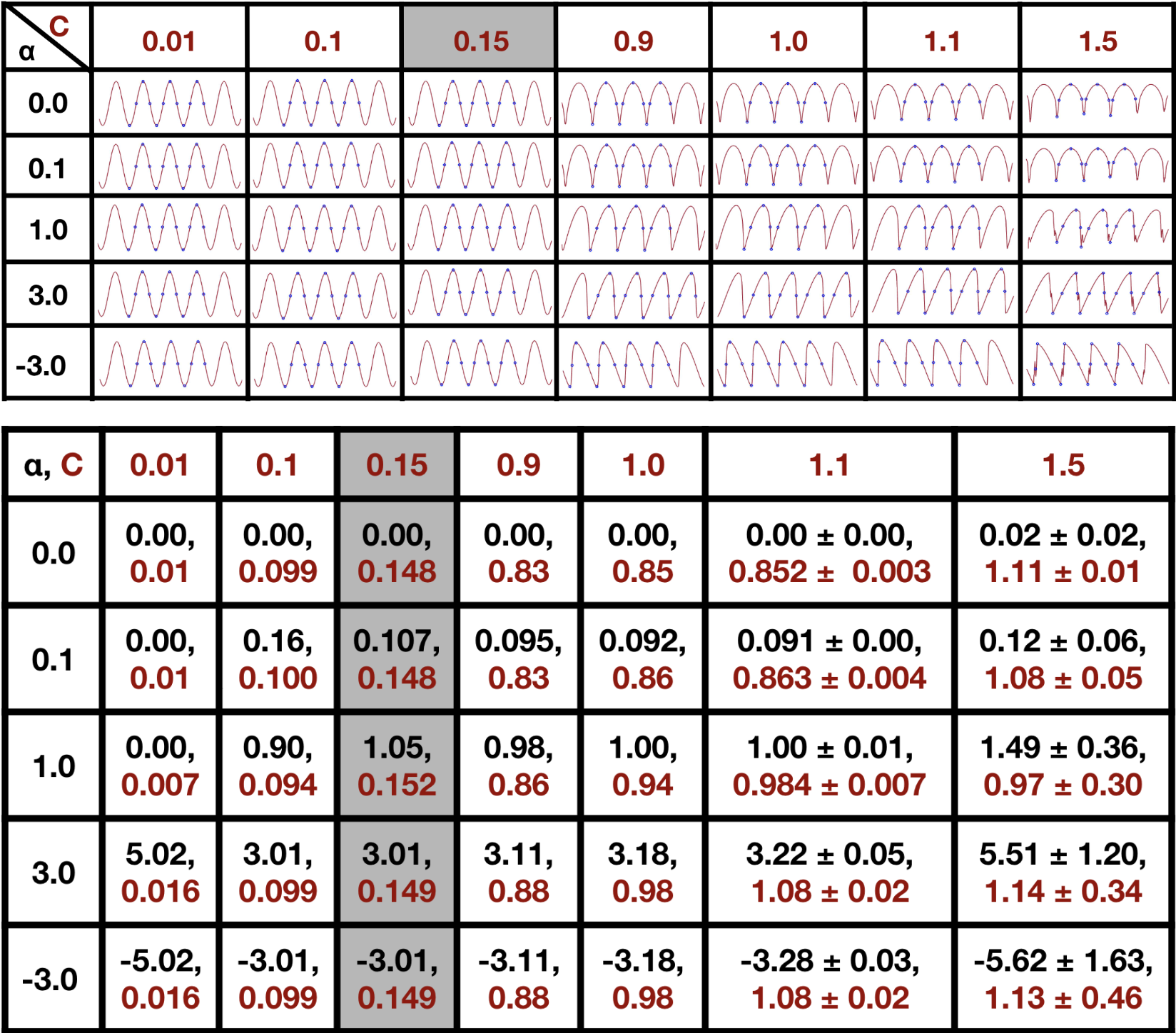}
    \caption{Series of simulated power interferogram fringes at various linewidth enhancement factors $\alpha$ and coupling constants $C$. Our analysis method is demonstrated on these simulated fringes by calculating a table of corresponding $\alpha$ and $C$ values. This analysis provides insight on the dependence of the asymmetry of fringes on the magnitude of the coupling constant and the magnitude and sign of the alpha factor. We have highlighted simulations with $C = 0.15$, which is close to the experimentally calculated coupling constant of our setup. We have also omitted uncertainties below 0.1\%.}
    \label{fig_alpha_factor_sim}
\end{figure}

In Fig.~\ref{fig_alpha_factor_sim}, we show an array of simulated fringes with varying $\alpha$ and $C$ values. For each set of fringes, we calculate the corresponding $\alpha$ and $C$ values as a demonstration of our analysis methods using Eq.~\ref{mean alpha} and error propagation similar to that in Eq.~\ref{error in alpha}. We notice high accuracy in the sign and magnitude of $\alpha$ and $C$ calculations, especially around what we found to be the experimentally realized coupling constant $C \approx 0.15$, which is well within the weak optical feedback regime. 

\textbf{Setup.} At each current supplied to the QCL, we aim to linearly sweep $\tau_{ext}$ while measuring the power emitted by our Fabry-Perot laser in the weak optical feedback regime of $C < 1$. Once we observe an interferogram under these conditions, we can calculate the LEF $\alpha$  and coupling constant $C$ using Eq.~\ref{mean alpha}. In practice, we must accommodate experimental nuances such as noise, sinusoidal $\tau_{ext}$ modulation, and an overall signal envelope. 

\begin{figure}
    \centering
    \includegraphics[width = .4\textwidth]{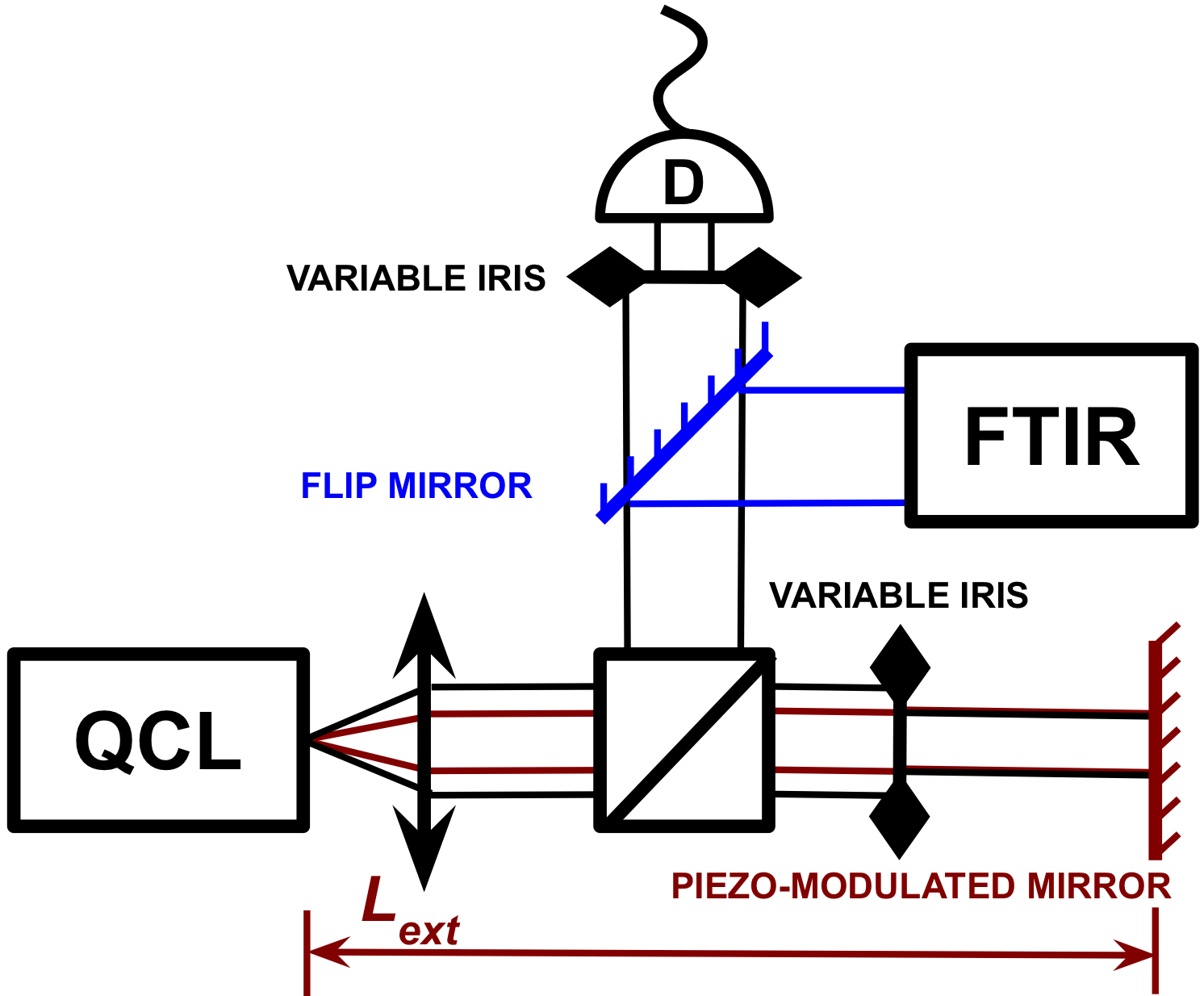}
    \caption{Setup used to measure the LEF of the laser. The maroon beam corresponds to optical feedback reflected off of an aluminum mirror pasted on a modulated piezoelectric actuator and attenuated by a variable iris and beamsplitter. By modulating the voltage across our piezo, we can modulate the external cavity length $L_{ext}$ and round trip time $\tau_{ext}$. Using the variable iris in front of our power detector, we can reduce optical feedback from back-reflection. The variable iris in front of the modulated mirror gives us control over the optical coupling constant $C$. Finally, we use a flip mirror and an FTIR to check that the Fabry-Perot laser remains single mode with and without optical feedback at every tested current value.}
    \label{fig_alpha_factor_setup}
\end{figure}

As shown in Fig.~\ref{fig_alpha_factor_setup}, we modulate the external cavity length $L_{ext}$ by placing our optical feedback mirror on a piezoelectric crystal that is driven at 145Hz. We place a variable iris in front of our feedback mirror to place our optical coupling in the weak feedback regime of $C \approx .15 < 1$. Using a 50/50 beamsplitter, we send part of our laser beam into a HgCdTe detector. Since our analysis assumes single-mode laser operation, we use a flip mirror to send the beam into a FTIR and verify that the Fabry-Perot laser remains in single-mode with optical feedback at every current value.

\begin{figure}
    \centering
    \includegraphics[width=.49\textwidth]{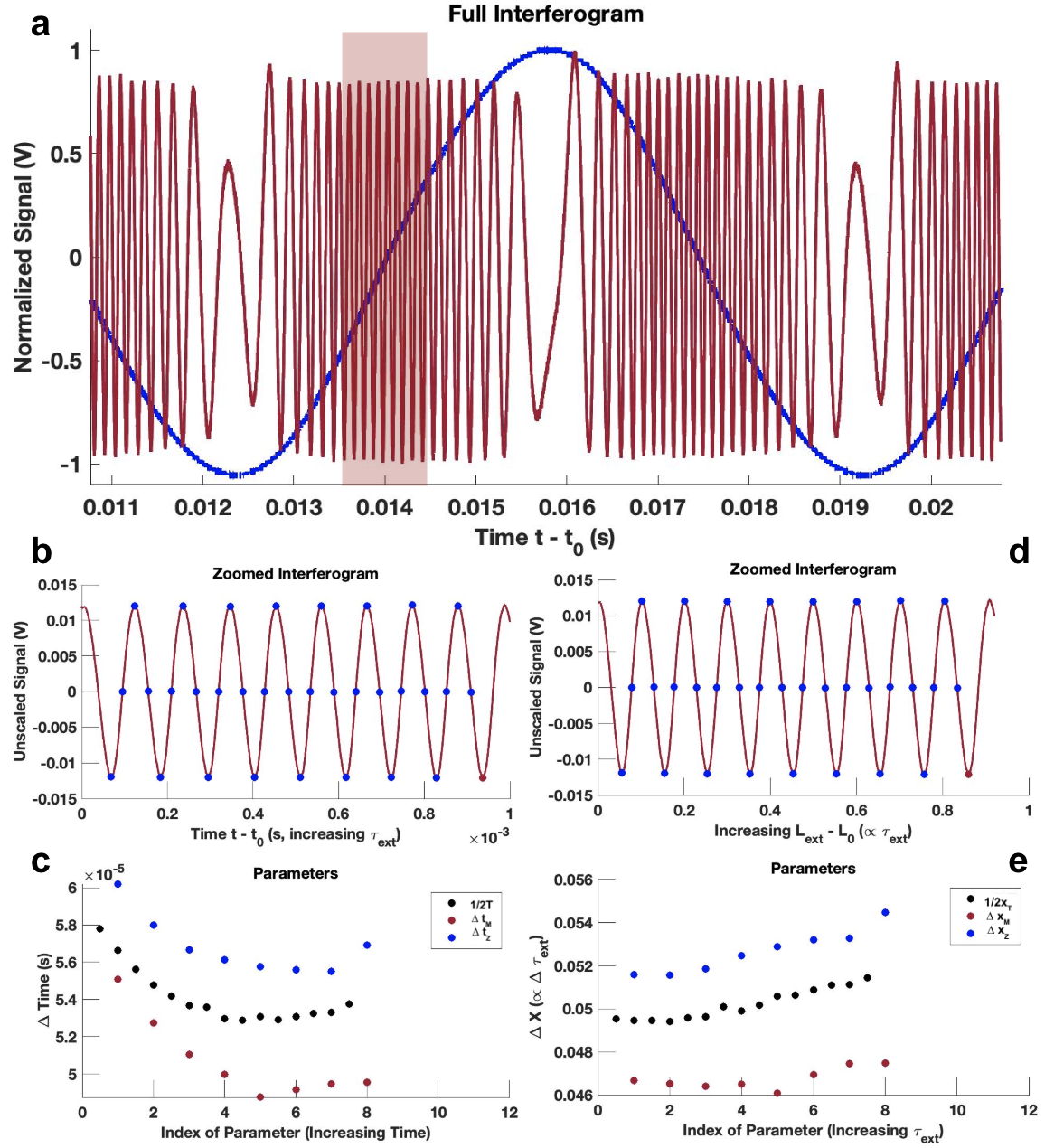}
    \caption{\textbf{a}, In maroon, we depict a period of averaged interferometric fringes in the emitted power of the QCL biased at 574.5 mA. These fringes are produced using a feedback mirror placed on a sinusoidally modulated piezoelectric. The flatness of the envelope of our fringes suggests that we are in the weak coupling regime of $C < 1$. The strain voltage of the piezoelectric, plotted in blue, tells us its expansion and thus the position of the feedback mirror versus time. For the purpose of our analysis, we must zoom into the highlighted, nearly linear portion of the sinusoidally driven feedback mirror. The larger the strain voltage the shorter the external cavity length $L_{ext}$ and external cavity round trip time $\tau_{ext}$. We must then flip the time-axis of the  highlighted fringes to plot them against a nearly linearly increasing $\tau_{ext}$ as shown in \textbf{b}. Notice the systematic curvature error in the period and other relevant analysis parameters, defined in Fig.~\ref{fig_alpha_factor_calculation}, of the fringes shown in \textbf{c}. This curvature arises from the inherently sinusoidal nature of the feedback mirror modulation, even around its inflection point. We then stretch the interferogram by plotting it against a cubic fit of the increasing position of the feedback mirror in \textbf{d}. The period of the fringes and other relevant analysis parameters exhibit a significantly flatter trend after this third-order correction.}
    \label{fig_stretched_interferogram}
\end{figure}

\textbf{Analysis.} Our analysis is predicated on interferometric fringes plotted against a linearly increasing $\tau_{ext}$. We thus stretch the measured interferogram as shown in Fig.~\ref{fig_stretched_interferogram}. In essence, we map the time coordinates $t$ of the interferogram data points to the corresponding position $x(t)$ of the feedback mirror taken from the strain of the piezoelectric crystal at that time, namely $(P(t), t) \mapsto (P(t), x(t))$. Since the change in the position $\Delta x(t)$ of the feedback mirror is linearly proportional to the change in the external cavity round-trip time $\Delta \tau_{ext}$, our analysis holds:  
\begin{equation}
    (\Delta t_M, \Delta t_Z, \Delta T) = (A\Delta x_M, A\Delta x_Z, A x_T)
\end{equation}
\begin{equation}\Rightarrow \overline{\alpha} = \frac{\overline{A\Delta x_M} - \frac{1}{2}\overline{A x_T}}{\overline{A\Delta x_Z} - \frac{1}{2}\overline{A x_T}} = \frac{\overline{\Delta x_M} - \frac{1}{2}\overline{ x_T}}{\overline{\Delta x_Z} - \frac{1}{2}\overline{ x_T}}
\end{equation}
where $A$ is a proportionality factor, and $\Delta x_M$, $\Delta x_Z$, and $x_T$ are defined analogously to the measurements in Fig.~\ref{fig_alpha_factor_calculation}. We also flip our interferogram appropriately so as to plot it against an increasing $\tau_{ext}$ as described in Fig.~\ref{fig_stretched_interferogram}. This allows us to accurately obtain the sign of the $\alpha$-factor of the QCL.

Finally, we calculate $\overline{\alpha} \pm \delta \alpha$ for currents swept at 0.5mA steps between lasing threshold at 561mA and 580mA as shown in Fig.~2c of the main text. For each current, we also verify the single mode operation of our laser light with and without optical feedback using a flip mirror as described in Fig.~\ref{fig_alpha_factor_setup}. This analysis used the definition $\alpha = \pdv{\chi_R}{N}/\pdv{\chi_i}{N}$, where ${\chi_R}$ and ${\chi_i}$ are the real and imaginary parts of the linear susceptibility;
however, in our derivation of the complex Ginzburg-Landau equation form of the QCL master equation we defined $\alpha = - \pdv{\chi_R}{N}/\pdv{\chi_i}{N}$. Thus the $\alpha$-factor relevant to the Ginzburg-Landau derivations is positive and stays around values $>1$ with a moderate increase in absolute value over the swept currents.

\newpage


%